\RequirePackage{fix-cm}
\documentclass[smallcondensed]{svjour3}     
\smartqed  
\usepackage{graphicx}
\usepackage{algpseudocode}
\usepackage{verbatim}
\usepackage{enumerate}
\usepackage{graphics}
\usepackage{multirow} 
\usepackage[center]{caption}
\usepackage{subfigure}
\usepackage{array}
\usepackage{algorithm}
\usepackage{inputenc}
\usepackage{array}
\usepackage{setspace} 
\usepackage{textcomp}
\usepackage{amsmath}
\usepackage{amsfonts}
\usepackage{amssymb}            
\usepackage{morefloats}
\usepackage{multirow}
\usepackage{nth}
\usepackage{xcolor}
\usepackage{hyperref}
\usepackage{bbm}

\newcommand{\uinc}{u^{\rm{inc}}}
\newcommand{\usca}{u^{\rm{sca}}}
\newcommand{\ii}{{\mathtt{i}}}
\newcommand{\dd}{{\,\text d}}
\newcommand{\kk}{{\kappa}}
\newcommand{\kr}{{\kappa r}}
\newcommand{\PP}{\mathbbm{P}}
\newcommand{\rmn}{{r^{\min}}}
\newcommand{\rmx}{{r^{\max}}}
\newcommand{\dun}{{\partial_{\nu}u}}
\newcommand{\bn}{{\bf n}}
\newcommand{\iim}{{\ii m}}

\newcommand{\eed}{\epsilon_{\text{dis}}} 
\newcommand{\eev}{\epsilon_{\text{evl}}} 

\newcommand{\CC}{\mathbbm{C}}
\newcommand{\NN}{\mathbbm{N}}
\newcommand{\RR}{\mathbbm{R}}

\newcommand{\ZZ}{{\bf Z}} 
\newcommand{\zz}{{\bf z}}
\newcommand{\Za}{{Z_1}}
\newcommand{\Zp}{{Z_p}}
\newcommand{\ZP}{{Z_P}}
\newcommand{\za}{{z_1}}
\newcommand{\zp}{{z_p}}
\newcommand{\zP}{{z_P}}

\newcommand{\Bcal}{\mathcal{B}} 
\newcommand{\Ccal}{\mathcal{C}} 
\newcommand{\Dcal}{\mathcal{D}} 
\newcommand{\Ecal}{\mathcal{E}} 
\newcommand{\Hcal}{\mathcal{H}} 
\newcommand{\Ical}{\mathcal{I}} 
 
\newcommand{\Lcal}{\mathcal{L}} 
\newcommand{\Rcal}{\mathcal{R}} 
\newcommand{\Ocal}{\mathcal{O}} 
\newcommand{\Scal}{\mathcal{S}} 
\newcommand{\Be}{{\RR}^2\setminus\overline{\Bcal}} 
\newcommand{\Et}{{\Ecal}^{\rm{tra}}} 
\newcommand{\Rt}{{\Rcal}^{\rm{tra}}} 
\newcommand{\SZ}{{\Scal(\ZZ)}}
\newcommand{\Sz}{{\Scal(z)}}
\newcommand{\IZ}{{\Ical_{\ZZ}}}
\newcommand{\IZr}{{\Ical_{Z_{\rr}}}}
\newcommand{\Sr}{{\Scal_{\rr}}}

\newcommand{\LS}{{\Lcal^2_{\!\dd\Scal}}}
\newcommand{\LG}{{\Lcal^2_{\!\dd F_{\ZZ}}}}

\DeclareMathOperator*{\argmin}{arg\,min}
\DeclareMathOperator*{\EE}{\mathbbm{E}}         
\DeclareMathOperator{\spnv}{span}

\newcommand{\Jm}{{\text J_m}}
\newcommand{\Hm}{{\text H_m}}
\newcommand{\jn}{{\text j_n}}

\newcommand{\aq}{{\theta}}                      
\newcommand{\al}{{\theta_{\ell}}}               
\newcommand{\aj}{{\varphi}}                     
\newcommand{\ab}{{z}}                           
\newcommand{\tr}{{\zeta}}                       

\newcommand{\xx}{{\bf x}}             
\newcommand{\rr}{{\bf r}}             
\newcommand{\rl}{{\bf r}_{\ell}}      
\newcommand{\ro}{{\boldsymbol\rho}}   
\newcommand{\rn}{{\bf\hat{r}}}              
\newcommand{\qn}{{\bf\hat{\boldsymbol\aq}}} 

\journalname{Journal of Scientific Computing}
\begin{document}

\title{Low-Dimensional Spatial Embedding Method for 
       Shape Uncertainty Quantification in Acoustic Scattering
       by 2D Star Shaped Obstacles
}

\titlerunning{Low-Dimensional Spatial Embedding Method for Shape UQ in Acoustic Scattering} 

\author{Yuval Harness 
}


\institute{Y. Harness \at
              Inria Bordeaux \\
              Tel.: +33-5-35-00-26-19\\
              Fax: +33-5-61-19-30-00\\
              \email{yuval.harness@inria.fr}           
}

\date{Received: date / Accepted: date}

\maketitle

\begin{abstract}
 This paper introduces a novel boundary integral approach of shape uncertainty quantification
 for the Helmholtz scattering problem in the framework of the so-called parametric method. 
 The key idea is to construct an integration grid whose associated
 weight function encompasses the irregularities and non-smoothness
 imposed by the random boundary. 
 Thus, the solution can be evaluated accurately with relatively low number of grid points.
 The integration grid is obtained by employing a low-dimensional
 spatial embedding using the coarea formula.
 The proposed method can handle large variation as well as non-smoothness 
 of the random boundary.
 For the ease of presentation the theory is restricted to star-shaped obstacles
 in low-dimensional setting.  
 Higher spatial and parametric dimensional cases are discussed, though, 
 not extensively explored in the current study.
\keywords{Uncertainty Quantification \and Shape Uncertainty \and Helmholtz 
    \and  Parametric Method \and Low-dimensional Embedding \and Coarea Formula} 
\subclass{MSC 49Q15 \and MSC 65C99 \and MSC 65N35 \and MSC 65R20 \and MSC 65Z05}
\end{abstract}

\section{Introduction}\label{sec:Section1}

 Considerable effort has been devoted in recent years to develop robust and 
 efficient computational strategies for the simulation of physical
 phenomena, that take into account shape uncertainty.
 Often, the problem is formulated as an elliptic \emph{partial differential equation} 
 (PDE) whose domain boundaries are uncertain.
 Such problems arise due to imperfections in manufacturing
 processes, e.g., in nano-optics where the production of nano particles 
 is, often, inaccurate relatively to nano-scale electromagnetic wave lengths \cite{bejan2000shape}.
 Other examples arise in the context of inverse problems, such as
 tomography where the visual representation of some hidden object is constructed
 by partial, and possibly noisy, measurements \cite{sethian1999level}.

 The common practice for quantifying uncertainty in computational models,
 is to employ the parametric method.
 The theoretical basis for the method was laid down by Wiener 
 \cite{gPC-Wiener-HomoChaos}. 
 The method itself was initially developed by Ghanem and Spanos \cite{gPC-GS-StochasticFE}, 
 and later generalized by Xiu and Karniadakis 
 \cite{gPC-WXK-FlowSimulations,gPC-XK-SteadyStateDiffusion,gPC-XK-Wiener-Askey}.
 In this approach the uncertain parameters are replaced by random quantities,
 and the problem is recast as a system with random input.
 The solution is estimated via global expansion of the random
 variables into a basis of uncorrelated functions.
 Thus, the stochastic problem is transformed into a deterministic 
 system in higher dimension. 
 The most popular expansions employed are 
 the \emph{Karhunen-Lo\`{e}ve expansion}, and the 
 \emph{generalized Polynomial Chaos} (gPC) expansion.

 The parametric approach, often, demonstrates
 superior performance in terms of computational effort over other 
 traditional methods, see \cite{gPC-Xiu-Book}
 for a detailed review. However, when the
 physical domain of the problem is uncertain the quantification
 by the parametric approach becomes much more challenging.
 The main difficulty stems from the fact that the problem
 is not characterized by smooth coefficients whose dependence
 on the random parameters is known. Thus, an accurate
 discretization which captures desired features of the solution
 for any realization of the random shape is not readily available.
 
 The stochastic collocation method and Monte Carlo sampling,
 which rely on samplings of the random parameters and the solution of 
 each realization deterministically are well established.
 However, for random domain problems each realization
 is, essentially, characterized by a different geometry and requires a custom
 discretization scheme. Generally, when the variations of the random domain
 are large and undergo complicated changes as a function of the random parameters, 
 these methods become extremely expensive to employ with
 prohibitive computational costs.

 To overcome the difficulties associated with the quantification
 of a random shape or domain, various techniques
 have been proposed. Typically, these are classified as one
 of the following: perturbation, fictitious domain, level-set  
 or random domain mapping. Perturbation techniques \cite{Shape-Perturbation}
 are straightforward and simple to apply, however, their applicability
 is limited to small shape deformations. The fictitious domain 
 \cite{Shape-Fictitious} and level-set methods \cite{Shape-LevelSet-Nouy20084663,Shape-LevelSet-XFEM}
 are based on embedding the random domain in a larger, 
 deterministic domain containing all possible realizations.
 These methods are capable of handling very irregular non-smooth
 geometries. However, the embedding introduces non-smoothness in the spatial region,
 that intersects with the random boundary. Thus, high-order convergence is 
 only partially ensured in the entire computational domain.

 The random domain mapping method \cite{Shape-TX2006b,Shape-TX2006a} 
 is the most common tool used for solving PDEs on uncertain domains. 
 The method is based on a realization-dependent coordinate transformation 
 uniformly mapping all the realizations of the domain to a fixed, reference configuration. 
 The variational formulation of the PDE on the random domain can then be 
 posed on the reference domain, reducing the problem to a
 PDE on a fixed domain with stochastic coefficients. 
 The transformed PDE whose domain is fixed is solved using standard
 techniques. However, the method is highly sensitive to the non-linear
 dependence of the problem on the random boundary. In case of complex
 evolution of the  shape, the random coefficients are difficult to obtain
 and typically exhibit highly varying behavior. 
 The common practice to overcome this difficulty is to impose 
 a highly accurate discretization grid, often combined with dimensionality reduction
 techniques, e.g., sparse grids, to ensure reasonable computational
 effort. See \cite{Shape-DomainMapping-Regularity,Shape-Harbrecht2016,Shape-Hptmr2015}
 for further details. 

 In this work an alternative method that attempts
 to mitigate the difficulties associated with the more standard techniques
 for PDEs on uncertain domains is proposed. The method is of boundary integral
 type \cite{BIE-CK-InverseScattering-Book} and, thus,
 can handle large shape deformations.
 The analysis is based on two observations. First, that as a function
 of the random boundary of the domain the solution is piece-wise smooth
 in the spatial domain. Second, that in practice we seek to
 approximate the outcome of a predetermined set of linear
 output functionals operating on the random boundary.
 The key idea is of this work is to construct an integration grid whose associated
 weight function encompasses the irregularities and non-smoothness
 imposed by the random boundary. Thus, the outcome of the functionals
 can be evaluated accurately with relatively low number of integration gridpoints.
 This idea is similar to certain classic numerical techniques for estimating
 integrals of highly oscillatory functions, which rely on
 oscillatory weighted Gaussian integration formulae.

 The proposed method constructs a discretization grid of the random surface
 for all possible realizations in two stages. In the first stage
 a spatial low-dimensional embedding of the family of random surfaces
 is constructed via the \emph{Coarea formula} \cite{Federer}. 
 The embedding, essentially, captures
 any irregular behavior of the random surface and a discretization
 is applied only on a compact region in the spatial domain.
 In the second stage a parametric grid corresponding
 to the low-dimensional spatial grid is imposed.
 A sparse or hierarchical parametric grid can be applied for dealing with 
 high dimensionality, while the spatial grid effectively ensures
 that the bulk variation of output functionals defined on the boundary
 is captured. In general, the method allows the handling of non-trivial geometries
 without the loss of accuracy in the region intersecting with the
 random interface.

 Since this is a first case study and for the ease of presentation,
 the discussion has been limited to time-harmonic wave scattering by 
 star-shaped obstacles. In the analysis and numerical study
 a $2D$ scattering object and low-dimensional parametric space 
 are assumed. More complicated examples in higher
 spatial and parametric dimensions are discussed.
 However, in-depth study of this topic is deferred to future work.
 For its simplicity, acoustic fluid-structure
 interaction has been chosen as the physical application. In that case,
 the solution represents small oscillations of pressure in a compressible
 ideal fluid. 
 The method and ideas presented in this work can
 also be applied to electrodynamics and elastodynamics.

 This work employs the null-field approach
 \cite{NFM-Martin-Book,NFM-Waterman-MatrixFormulation,NFM-Waterman-T-Matrix,NFM-Wriedt-Review},
 which in contrary to the better known \emph{boundary element method} (BEM) 
 \cite{BIE-sauter2010boundary} and the Nystr\"{o}m method 
 \cite{kress2013linear}, does not involve singular integrals.
 Null-field methods are fast and much easier to implement compared to 
 BEM and the Nystr\"{o}m method.
 Their applicability range is, however, more limited.
 The null-field reconstruction technique 
 \cite{NFM-EM-Random-Shape-1D,NFM-Optimal-Reconstruction}
 is inherently stable, admits a-priori error evaluation, and facilitates the extraction
 of features of interest without prior estimation of the entire solution.
 The method enables us to perform analysis from a purely geometric point of view, 
 which avoids the additional complications associated with integration
 of weakly singular kernels.
 Combining low-dimensional surface embedding with BEM and Nystr\"{o}m method 
 can be foreseen in a future study.

 The paper is organized as follows. The fundamentals 
 of the null-field reconstruction method for the time-harmonic wave scattering
 problem is presented in Section \ref{sec:Section2}. 
 Section \ref{sec:Section3} reviews the procedure of optimal 
 reconstruction from a numerical linear algebra point of view. 
 Section \ref{sec:Section4} consists of the main theoretical results of this work
 and includes the formulation of the problem.
 In Section \ref{sec:Section5} the proposed method is applied to 
 a class of randomly shaped polygonal cylinders, as a proof of
 concept that the suggested method can, indeed, handle complex non-smooth shapes.
 Summary of the results, conclusions, and suggestions for applying the method 
 in more complicated scenarios are given in Section \ref{sec:Section6}.

\section{Null-Field Reconstruction for Time Harmonic Wave Scattering}\label{sec:Section2}

 In this section a brief review on the null-field
 reconstruction method for the time-harmonic wave scattering problem is given.
 The time-harmonic acoustic scattering problem is presented,
 followed by a review of the fundamental theory of null-field methods.
 The main idea of the null-field reconstruction 
 technique for time-harmonic wave scattering is presented in the concluding
 subsection.

 \subsection{Acoustic Scattering by Impenetrable Obstacles}

 Let $\Bcal$ denote a bounded domain in $\RR^2$ representing an
 impenetrable obstacle with boundary $\Scal$.
 We denote by $\overline{\Bcal}=\Bcal\cup\Scal$ the closure of $\Bcal$.
 Let $\Be$ be the unbounded exterior region occupied by a uniform
 medium.
 Let $\rr\in\RR^2$ denote a general spatial point,
 \[
  \rr
   = 
  (r\cos\aq,r\sin\aq)
   \,,
 \]
 For an incident time-harmonic field
 ${\uinc}(\rr)$ 'illuminating' the obstacle, the scattered field
 ${\usca}(\rr)$ satisfies the following exterior boundary value problem:
 \begin{equation}\label{def:sec2_Helmholtz}
  \Delta {\usca}+\kappa^2{\usca} = 0 \;\;\; \forall\,\rr\in \Be
  \,,
 \end{equation}
 \begin{equation}\label{def:sec2_SoundSoftHard}
  {\usca}=-{\uinc} \;\text{or}\; \partial_{\nu}{\usca}=-\partial_{\nu}{\uinc} \;\;\; \text{on} \;\;\; \Scal
  \,,
 \end{equation}
 \begin{equation}\label{def:sec2_Sommerfeld}
  r^{1/2}\left(\partial_r{\usca}-\ii\kappa {\usca}\right) \rightarrow 0 \;\;\; \text{as} \;\;\; r\rightarrow\infty
  \,,
 \end{equation}
 where $\ii=\sqrt{-1}$.
 Equation (\ref{def:sec2_Helmholtz}) is known as the \emph{Helmholtz} equation,
 where $\Delta$ is the Laplacian and $\kappa$ is the \emph{wavenumber}.
 Equation (\ref{def:sec2_SoundSoftHard}) specifies the boundary
 condition, depending on the physical problem: adopting the acoustic
 terminology, it is \emph{sound-soft} for Dirichlet problems and
 \emph{sound-hard} for Neumann problems.
 Here $\nu$ is the unit outward normal to $\Scal$
 and $\partial_{\nu}u$ is the normal derivative of $u$.
 The last condition (\ref{def:sec2_Sommerfeld}),
 known as the \emph{Sommerfeld radiation condition},
 ensures that the scattered field propagates from
 the obstacle to infinity.
 The solution of the exterior scattering problem is unique.
 A solution to the Helmholtz equation is called a \emph{wavefunction}.
 A wavefunction satisfying the Sommerfeld condition
 (\ref{def:sec2_Sommerfeld}) is called an \emph{outgoing}
 wavefunction.

 \subsection{Null-Field Theory Fundamentals}

 Null-field methods for the acoustic scattering problem
 (\ref{def:sec2_Helmholtz}) are based on Green's second
 theorem
 \begin{equation}\label{def:sec2_Green2nd}
 \iint_{\Bcal}\left(
  w\Delta v - v\Delta w
 \right)\dd V
  =
 \int_{\Scal}\left(
   w\frac{\partial v}{\partial\nu}
   -
   v\frac{\partial w}{\partial\nu}
 \right)\dd\Scal
 \,,
 \end{equation}
 which holds for any bounded domain $\Rcal$ with
 a Lipschitz piecewise smooth boundary $\Scal$,
 where $v$ and $w$ are scalar fields,
 and ${\partial v}/{\partial\nu}$ and ${\partial w}/{\partial\nu}$
 denote corresponding normal derivatives.

 Let ${\psi}$ be an outgoing wavefunction
 and let $u$ denote the total field, ${\uinc}+{\usca}$.
 Assuming ${\psi}$ is analytic in $\Be$, it can be shown by
 (\ref{def:sec2_Green2nd}) that
 \[
 \int_{\Scal}\left(
  u\frac{\partial {\psi}}{\partial\nu}
   -
  {\psi}\frac{\partial u}{\partial\nu}
 \right)\dd\Scal
  =
 \int_{\Scal}\left(
  {\uinc}\frac{\partial {\psi}}{\partial\nu}
   -
  {\psi}\frac{\partial {\uinc}}{\partial\nu}
 \right)\dd\Scal
 \,.
 \]
 Thus, for a sound-soft obstacle ($u=0$ on $\Scal$)
 \begin{equation}\label{def:sec2_SoftNullField}
 \int_{\Scal}{\psi}\frac{\partial u}{\partial\nu}\dd\Scal
  =
 -\int_{\Scal}\left(
  {\uinc}\frac{\partial {\psi}}{\partial\nu}
   -
  {\psi}\frac{\partial {\uinc}}{\partial\nu}
 \right)\dd\Scal
 \,,
 \end{equation}
 while for a sound-hard obstacle  ($\dun=0$ on $\Scal$)
 \begin{equation}\label{def:sec2_HardNullField}
 \int_{\Scal}u\frac{\partial {\psi}}{\partial\nu}\dd\Scal
  =
 \int_{\Scal}\left(
  {\uinc}\frac{\partial {\psi}}{\partial\nu}
   -
  {\psi}\frac{\partial {\uinc}}{\partial\nu}
 \right)\dd\Scal
 \,.
 \end{equation}
 Using (\ref{def:sec2_SoftNullField})
 or (\ref{def:sec2_HardNullField}), an infinite set of 
 equations can be produced from which $u$ or $\dun$ on
 $\Scal$ are approximated. In practice, one chooses
 a finite subset of equations of the form of
 (\ref{def:sec2_SoftNullField}) and (\ref{def:sec2_HardNullField})
 which are employed to optimally reconstruct the scattered
 field without an explicit estimation of $u$ or $\dun$ on
 $\Scal$. The core idea of reconstruction by functionals is 
 presented in the next subsection, while the numerical procedure 
 for its practical implementation is covered in 
 Section \ref{sec:Section3}.

 \subsection{Reconstruction of Surface Functionals}

 Typically we are interested in estimating features of interest
 which are expressed by the unknown surface density, $u$ or $\dun$ on $\Scal$.
 Often such features are the outcomes of functionals in an
 appropriate \emph{Hilbert space}.
 Indeed, let $h$ denote the complex conjugate of
 $u$ or $\dun$ on $\Scal$. Then for a general non-smooth surface
 $\Scal$, the surface density $h$ belongs to the complex Hilbert space $\LS(\Scal)$
 whose inner-product and norm are defined by
 \[
  \left< g\,,h\right>_{\LS}=\int_{\Scal}h^*(\rr)g(\rr)\dd\Scal
   \,, \;\;\;
  \left\|h\right\|_{\LS}=\sqrt{\left< h\,,h\right>_{\LS}}
   \,,
 \]
 where $h^*$ denotes the complex conjugate of $h$ and $\dd\Scal$ is the 
 induced volume form on the surface.
 Recall that by \emph{Riesz representation theorem} any bounded linear
 functional $\LS\rightarrow\CC$ operating on surface densities,
 is of the form $\left< f\,,h\right>_{\LS}$. We call such
 functionals {\em surface functionals}.

 In this work we focus on the estimation of the \emph{scattering coefficients}
 of the expansion of ${\usca}$ to cylinder harmonics in
 the $2D$ case.
 These coefficients, denoted by $b_m$, satisfy
 \[
  {\usca}(\rr) =
   \sum_{m=-\infty}^{\infty}b_m\Hm(\kr)e^{\ii m\aq}
   \,,
 \]
 where $\Hm(z)$ denotes the $m$th-order \emph{Hankel function of the first kind}.
 See \cite{BIE-CK-InverseScattering-Book} for further details.
 The scattering coefficients are very useful features of the
 surface density, since they can easily express other important 
 quantities such as the \emph{far-field pattern} and the 
 \emph{radar cross section} \cite{NFM-Thesis}.

 Consider the sound-soft case (\ref{def:sec2_SoftNullField}).
 Using the Hilbert space notation, it follows that the
 scattering coefficients satisfy
 \begin{equation}\label{def:sec2_SoftScatterCoefInnerProduct2D}
  \frac{\ii}{4}
  \left<
   \Jm(\kr)e^{-\iim\aq}
  \,,\,
   \partial_{\nu}u^*(\rr)
  \right>_{\LS}
   = b_m
  \,, \;\;\; \;\;\; \forall\;m\in\ZZ
  \,,
 \end{equation}
 where $\Jm(z)$ denotes the $m$th-order
 \emph{Bessel function of the first kind},
 and $\partial_{\nu}u^*(\rr)$ denotes the complex conjugate
 of the surface density $\partial_{\nu}u(\rr)$.
 In practice only $\{b_m\}_{|m|\leq\mu}$ satisfying
 \begin{equation}\label{def:sec2_BesselSeriesTruncationParameter}
  \mu=\lceil3\kappa\rmx\rceil
   \,, \;\;\;
  \rmx
   =
  \inf\left\{
   \left|\rr\right| \,\left|\, \rr\in\Be \right.
  \right\}
   \,,
 \end{equation}
 are required for an accurate description of $\usca$,
 see \cite{NFM-Optimal-Reconstruction} for more details.
 Similar expressions can be derived for the sound-hard case.

 The core idea of the reconstruction procedure is to approximate
 the outcome of target functionals
 (\ref{def:sec2_SoftScatterCoefInnerProduct2D}) and
 without producing an explicit approximation of the
 surface density $\dun$ on $\Scal$. This, generally, allows us to
 handle complex geometries as well as irregular
 or singular surface densities much more accurately . 
 Explicitly, we approximate the each outcome
 (\ref{def:sec2_SoftScatterCoefInnerProduct2D})
 by a linear combination of the following form
 \[
  \frac{\ii}{4}
  \left<
   \Jm(\kr)e^{-\iim\aq}
  \,,\,
   \overline{\partial_{\nu}u}
  \right>_{\LS}
   \approx
  \left<
   \sum c_{m,l}{\psi}_{\ell}
  \,,\,
   \overline{\partial_{\nu}u}
  \right>_{\LS}
 \,,
 \]
 where $\{{\psi}_{\ell}\}$ 
 are predetermined sets of functionals,
 whose outputs are either known or can be calculated directly.
 We call such functionals the \emph{information functionals}.

 As shown in (\ref{def:sec2_SoftNullField}) the outcome of the 
 information functionals are readily available if ${\psi}_{\ell}$ 
 are outgoing wavefunctions whose 
 singularities are located in $\overline{\Bcal}$. 
 Hence, given the information
 \[
  \left<{\psi}_{\ell}\,,\,\overline{\partial_{\nu}u}\right>_{\LS}
   =
  a_{\ell}
  \,,
 \]
 the outcome of the target functional can be approximate by
 \[
  \frac{\ii}{4}
  \left<
   \Jm(\kr)e^{-\iim\aq}
  \,,\,
   \overline{\partial_{\nu}u}
  \right>_{\LS}
   \approx
  \sum c_{m,l}a_{\ell}
 \,, \;\;\;\forall\;m\in\ZZ \,.
 \]
 Obtaining the coefficients $\{c_{m,\ell}\}$ while ensuring measurable
 error bounds of the estimations of the scattering coefficients
 can be achieved by \emph{reconstruction kernel} approximation
 which is the main topic of Section (\ref{sec:Section3}).

\section{Optimal Reconstruction with A-priori Error Estimate}\label{sec:Section3}

 In this section we review the procedure for the recovery of target functionals by information functionals
 in general Hilbert space. 
 Reconstruction problems often involve regularization parameters
 which govern the stability and accuracy of the procedure. The
 result of the optimization of the reconstruction with respect to the
 regularization parameters is referred to as optimal reconstruction.
 Optimal reconstruction can be traced back to the notion of optimal recovery
 \cite{OR-GW-OptimalApproximation,OR-MR-OptimalEstimation}.
 A more modern analysis from an inverse problem point of
 view can be found in
 \cite{Moment-Louis-FeatureReconstruction,Moment-Louis-UnifiedApproach,Moment-LM-MollifierMethod}.

 We begin with the definition of the reconstruction problem and the notion of 
 reconstruction kernel. 
 This is followed by a brief description of the numerical procedure including error analysis. 
 The final part elaborates on proper numerical integration rules,  
 that are needed for the error estimates.
 The method and error analysis presented here, as well as further technical
 details have been initially introduced in \cite{NFM-Optimal-Reconstruction}.

 \subsection{The Reconstruction Problem and Reconstruction Kernels}

 Let $\Hcal$ be a complex Hilbert space, whose inner-product is denoted
 by $\left<\,,\,\right>_{\Hcal}$.
 The reconstruction problem is to approximate
 a finite set of target functionals
 \begin{equation}\label{def:sec3_TargetFunctionals}
  \left< g_m\,,h\right>_{\Hcal}=b_m
   \,, \;\;\;
  g_m\in\Hcal
   \,, \;\;\;
  m=1,2,{\ldots},M
   \,,
 \end{equation}
 by a given finite set of information functionals,
 \begin{equation}\label{def:sec3_InformationFunctionals}
  \left< f_{\ell}\,,h\right>_{\Hcal}=a_{\ell}
   \,, \;\;\;
  f_{\ell}\in\Hcal
   \,, \;\;\;
  l=1,2,{\ldots},L
   \,,
 \end{equation}
 where the element $h\in\Hcal$ is unknown.

 \begin{definition}
 Let $\left\|\,\,\,\right\|_{\Hcal}$ denote the norm induced by
 $\left<\;,\,\right>_{\Hcal}$ in $\Hcal$, and let $\Ccal_m^L$ denote a
 closed convex subset of $\CC^L$.
 A linear combination
 $\sum_{\ell=1}^{L}\widehat{c}_{m,\ell}f_{\ell}$
 whose coefficients
 $(\widehat{c}_{m,1},{\ldots},\widehat{c}_{m,L})\subset\Ccal_m^L$
 satisfy the minimality condition
 \begin{equation}\label{def:sec3_OptimalReconstruction}
  \left\|
   g_m-\sum_{\ell=1}^{L}\widehat{c}_{m,\ell}f_{\ell}
  \right\|_{\Hcal}
   \leq
  \left\|
   g_m-\sum_{\ell=1}^{L}c_{m,\ell}f_{\ell}
  \right\|_{\Hcal}
   \;\;\;
  \forall
  \left(c_{m,1},{\ldots},c_{m,L}\right)\in\Ccal_m^L
   \,,
 \end{equation}
 is called an \emph{optimal reconstruction kernel} of the target functional $g_m$
 by the information functionals $\{f_{\ell}\}$ over $\Ccal_m^L$.
 \end{definition}
 \begin{remark}
  Clearly, (\ref{def:sec3_OptimalReconstruction})
  is a projection on a convex set. The key point which is addressed later,
  is how to determine the convex set $\Ccal_m^L$. Note that almost
  no prior knowledge on the element $h$ is assumed.
 \end{remark}

 We will show in the next subsection, that obtaining 
 (\ref{def:sec3_OptimalReconstruction}) vastly exceeds our needs. 
 In practice, it is sufficient to obtain an approximation satisfying
 \begin{equation}\label{def:sec3_ReconstructionKernel}
  \left\|
   g_m-\sum_{\ell=1}^{L}\widehat{c}_{m,\ell}f_{\ell}
  \right\|_{\Hcal}
   \leq
  \epsilon\left\| g_m  \right\|_{\Hcal}
   \;\;\;
  \left(\widehat{c}_{m,1},{\ldots},\widehat{c}_{m,L}\right)\in\Ccal_m^L
   \,,
 \end{equation}
 with respect to some predetermined threshold, $\epsilon>0$.
 In that case the linear combination $\sum_{\ell=1}^{L}\widehat{c}_{m,\ell}f_{\ell}$
 is simply called a \emph{reconstruction kernel} (i.e., not optimal).

 \subsection{The Discrete Reconstruction Procedure with Error Analysis}

 For the evaluation of the reconstruction kernel,
 we assume a finite dimensional discretization satisfying 
 the following definition.
 \begin{definition}\label{def:sec3_InnerProductPreservingDiscretization}
  Let $\Hcal_B$ be a bounded subset of a Hilbert space $\Hcal$
  whose inner-product is denoted by $\left<\;,\,\right>_{\Hcal}$.
  A mapping,
  \[
   T_{\Hcal_B}:\Hcal_B\rightarrow\CC^{1\times P}
    \,, \;\;\;
   P\in\NN
    \,,
  \]
  is called an inner-product preserving discretization of $\Hcal_B$ of
  accuracy $\eed>0$ if
  \begin{equation}\label{def:sec3_InnerProductPreserve}
   \left|
    \overrightarrow{f}\overrightarrow{g}^*
   -
    \left< f,g\right>_{\Hcal}
   \right|
    <
   \eed
    \,, \;\;\;
   \forall\,f,g\in\Hcal_B
    \,,
  \end{equation}
  where $\overrightarrow{f}=T_{\Hcal_B}(f)$ and
  $\overrightarrow{g}=T_{\Hcal_B}(g)$.
  The vectors $\overrightarrow{f}$ and
  $\overrightarrow{g}$ are called the corresponding
  inner-product preserving discretizations of
  $f$ and $g$ on $\Hcal_B$.
 \end{definition}

 Let $\overrightarrow{f_{\ell}},\overrightarrow{g_m}\in\CC^{1\times P}$
 denote inner-product preserving discretizations of some $f_{\ell},g_{\ell}\in\Hcal_B$,
 respectively.
 Let $\eta\in{\Hcal}$ denote the orthogonal projection of $h$ on the
 subspace spanned by $\Hcal_B$. 
 By definition (\ref{def:sec3_InnerProductPreservingDiscretization}) we obtain
 \[
  \left|\overrightarrow{f_{\ell}}\overrightarrow{\eta}^*-\left< f_{\ell}\,,\,h\right>_{\Hcal}\right|
   \,,
  \left|\overrightarrow{g_m}\overrightarrow{\eta}^*-\left< g_m\,,\,h\right>_{\Hcal}\right|
  \leq \|\eta\|\eed
   \,,
 \]
 for all $\ell\in\{1,2,{\ldots},L\}$ and $m\in\{1,2,{\ldots},M\}$.
 Hence, given the information (\ref{def:sec3_InformationFunctionals})
 and an approximation of $\overrightarrow{g_m}$,
 \begin{equation}\label{asm:sec3_TargetFunctionalApproximation}
  \widehat{g}_m
   =
  \sum_{\ell=1}^{L}\widehat{c}_{m,\ell}\overrightarrow{f_{\ell}}
   \,,
 \end{equation}
 we can reconstruct the unknown target coefficients (\ref{def:sec3_TargetFunctionals})
 via
 \begin{equation}\label{asm:sec3_TargetCoefficientReconstruction}
  b_m
   =
  \left< g_m\,,\,h\right>_{\Hcal}
   \approx
  \widehat{g}_m\overrightarrow{\eta}^*
   =
  \sum_{\ell=1}^{L}\widehat{c}_{m,\ell}\overrightarrow{f_{\ell}}\overrightarrow{\eta}^*
   \approx
  \sum_{\ell=1}^{L}\widehat{c}_{m,\ell}a_{\ell}
   \,.
 \end{equation}

 To evaluate the error of the reconstruction (\ref{asm:sec3_TargetCoefficientReconstruction})
 we denote for each $\ell=1,2,\ldots,L$ and $m=1,2,\ldots,M$ the discretization errors
 \[
  \epsilon_{\ell}=\left< f_{\ell}\,,h\right>_{\Hcal}
                 -
                  \overrightarrow{f_{\ell}}\overrightarrow{\eta}^*
   \,,
  \delta_m=\left< g_m\,,h\right>_{\Hcal}
          -
          \overrightarrow{g_m}\overrightarrow{\eta}^*
   \,,
 \]
 and obtain the following estimate
 \[
  b_m
   -
  \sum_{\ell=1}^{L}\widehat{c}_{m,\ell}a_{\ell}
   =
  \delta_m
   +
  (\overrightarrow{g_m}-\widehat{g}_m)\overrightarrow{\eta}^*
   +
  \sum_{\ell=1}^L\widehat{c}_{m,\ell}\epsilon_{\ell}
   \,,
 \]
 where our assumption ensure that $|\delta_m|,|\epsilon_{\ell}|\leq \|\eta\|\eed$.
 Note that $(\overrightarrow{g_m}-\widehat{g}_m)\overrightarrow{\eta}^*$ is
 the projection error which can not be reduced if the set of information
 functionals, $\{f_{\ell}\}$, is predetermined.

 To control the error we impose the following regularization constraint
 \begin{equation}\label{asm:sec3_RegularizationRule}
  |\widehat{c}_{m,\ell}|
   \leq
  \eev / \eed
   \,, \;\;\;
  \eev\geq\eed
   \,,
 \end{equation}
 where $\eev$ is a chosen or given evaluation error bound.
 Thus, we obtain 
 \begin{equation}\label{def:sec3_ReconstructionError}
  \left| b_m - \sum_{\ell=1}^{L}\widehat{c}_{m,\ell}a_{\ell} \right|
   \leq
  \|\eta\|\cdot\eed + \left|(\overrightarrow{g_m}-\widehat{g}_k)\overrightarrow{\eta}^*\right|
   + L\cdot \|\eta\|\cdot\eev
   \,.
 \end{equation}
 The regularization constraint (\ref{asm:sec3_RegularizationRule})
 explicitly defines the convex set $\Ccal_m^L$ in 
 (\ref{def:sec3_OptimalReconstruction}) as
 \[
  \Ccal_m^L
   =
  \left\{
   (c_1,c_2,\ldots,c_L)
    \, \left| \,
   |c_{\ell}|\leq\frac{\eev}{\eed} \right.
  \right\}
   \,.
 \]

 The error estimate (\ref{def:sec3_ReconstructionError}) implies
 that it is sufficient to obtain an approximation (\ref{asm:sec3_TargetFunctionalApproximation})
 satisfying $\left\|(\overrightarrow{g_m}-\widehat{g}_k)\right\|\leq\eev$.
 Indeed, in that case (\ref{def:sec3_ReconstructionError}) reduces to
 \[
  \left| b_m - \sum_{\ell=1}^{L}\widehat{c}_{m,\ell}a_{\ell} \right|
   \leq
   \|\eta\|\cdot\eed 
    +
   \left(L+1\right)\cdot \|\eta\|\cdot\eev
   \,.
 \]
 Often, the summation of evaluation errors $\sum_{\ell=1}^L\widehat{c}_{m,\ell}\epsilon_{\ell}$ 
 is not cumulative. Thus, the overall error is typically $\Ocal\left(\eev\right)$
 assuming $\|\eta\|=\Ocal(1)$.
 This facilitates an efficient approximation technique which is
 based performing successive \emph{singular value decompositions} on
 subsets of information functionals. The technique
 was presented in \cite{NFM-Optimal-Reconstruction} and demonstrated
 high stability and good convergence properties. Further details 
 including different variants of the technique can be found in \cite{NFM-Thesis}.


 \subsection{Inner-Product Preserving Discretization and Numerical Integration}

 Obtaining inner-product preserving discretizations of surface functionals
 is a fundamental issue. Let us focus on the case, as in this work,
 where $\{f_{\ell}\}$ and $\{g_m\}$ are smooth functions in
 some $\Hcal=\Lcal^2$ space with an inner-product, 
 \begin{equation}\label{def:sec3_InnerProduct}
  \left< f\,,\,g\right>_{\Hcal}
   =
  \int_{\Dcal}f g^*\,\omega\dd{s}
   \,,
 \end{equation}
 where $\Dcal\subset\RR^d$ is compact and Jordan measurable,
 $g^*$ is the complex conjugate of $g$
 and $\omega$ is a proper \emph{weight function}.

 To numerically compute the integrals \ref{def:sec3_InnerProduct},
 we observe that it is sufficient to employ an integration
 rule which is accurate on the finite dimensional 
 subspace of smooth functions spanned by $\{f_{\ell}\}\cup\{g_m\}$.
 Thus, we assume the availability of a standard rule of the following form
 \[
  \left< f,g\right>_{\Hcal}
   \approx
  \sum_{i=1}^Nf(\sigma^{(i)})g^{*}(\sigma^{(i)})\omega^{(i)}
   \;\;\;
  \forall\,f,g\in\spnv\left(\{f_{\ell}\}\cup\{g_m\}\right)
   \,,
 \]
 with \emph{integration nodes} $\sigma^{(1)},{\ldots},\sigma^{(N)}$
 contained in $\Dcal$ and real positive
 weights $\omega^{(1)},{\ldots},\omega^{(N)}$.
 Discretizing an element $f\in\Hcal$ as a weighted
 \emph{gridfunction}
 \begin{equation}\label{def:sec3_WeightedGridfunction}
  \overrightarrow{f}=\left(f(\sigma^{(1)})\sqrt{\omega^{(1)}},{\ldots},f(\sigma^{(N)})\sqrt{\omega^{(N)}}\right)
  \,,
 \end{equation}
 essentially, satisfies the inner-product preserving assumption
 (\ref{def:sec3_InnerProductPreserve}) if $N\in\NN$ is sufficiently large.
 The number of elements $N$ required for an effective
 inner-product preserving discretization depends on the convergence
 rate of the numerical integration formula and, typically, under 
 some smoothness assumption of the integrands.
 Indeed, if the weight function $\omega$ encompasses all the singularities while
 $f(s)$ and $g(s)$ are analytic, a \emph{Gaussian} numerical integration rule with respect to $\omega$
 ensures exponential convergence. Note that the weights of
 Gaussian rules are always positive and uniformly bounded. See
 \cite{davis2007methods} for more details.

\section{Surface Embedding of $2D$ Random Star-Shaped Obstacles}\label{sec:Section4}

 In this section the main theoretical contribution of this paper is presented.
 The first two subsections cover the setting of the problem, where
 Subsection (\ref{sec:SubSection4_1}) defines the random shape properties,
 and Subsection (\ref{sec:SubSection4_2}) covers relevant components of the 
 \emph{generalized Polynomial Chaos} (gPC) expansion theory.
 The chosen framework leads to a reconstruction problem in a Hilbert space.
 A concise discussion on the disadvantages of naive discretization of 
 the reconstruction problem concludes Subsection (\ref{sec:SubSection4_2}).
 
 In Subsection (\ref{sec:SubSection4_3}) we present an analytic approach
 for overcoming the difficulties associated with the naive discretization approach.
 Using the Coarea formula we construct a low-dimensional spatial embedding within the 
 family of random surfaces, which facilitates a natural choice for setting 
 a cubature rule in a compact region of $\RR^2$.
 The chosen integration weight function is a strictly positive minimal variance 
 quantity encompassing the irregularities of the family of random surfaces.

 In Subsections (\ref{sec:SubSection4_4}) and (\ref{sec:SubSection4_5})
 we focus on the case of a single
 random variable describing the randomness of the object. Using the
 \emph{implicit function theorem} we obtain explicit formulas including
 full characterization of the singular behavior of the integration weight function.
 The usage of the single random variable formulation as a building block
 for the more general case of multiple random variables is considered
 and discussed in Section (\ref{sec:Section6}).

 Subsections (\ref{sec:SubSection4_6}) and (\ref{sec:SubSection4_7})
 are devoted to the demonstration of the preceding theoretical parts 
 on a model problem of a randomly oriented elliptic cylinder.
 The random orientation problem is a very simple 'toy' problem. However,
 it allows us us to demonstrate in an affable fashion
 the implementation of the theory.

 \subsection{The $2D$ Random Shape Setting}\label{sec:SubSection4_1}

 For brevity, we focus on the sound-soft case and assume that $\Bcal$ 
 represents a star-shaped obstacle in $\RR^2$ whose boundary, $\Scal$, 
 depends smoothly on a real valued vector of mutually independent and 
 continuous random variables
 \[
  \ZZ=(\Za,{\ldots},\ZP)
   \,, \;\;\;
  P\in\NN
   \,.
 \]
 The boundary $\Scal$ is, however, not assumed to be
 uniformly smooth in the spatial domain.
 We assume that each random variable $\Zp$ has finite even moments
 \begin{equation}\label{asm:sec4_FiniteStatMoments}
  \EE\left[\Zp^{2n}\right]
   =
  \int_{\Ical_{\Zp}}z_p^{2n}\frac{dF_{\Zp}}{d\zp}\dd\zp
   <
  \infty
   \,,\;\;\;
  n\in\left\{0,1,{\ldots},N\right\}
   \,,
 \end{equation}
 where $\Ical_{\Zp}$ is the support of $\Zp$
 and $\frac{dF_{\Zp}(\zp)}{d\zp}$ is the \emph{probability density function} of $\Zp$. 
 Property (\ref{asm:sec4_FiniteStatMoments}) effectively ensures the existence of 
 surface functionals suitable for the reconstruction of the scattering coefficients.
 
\begin{figure}[!h]
  \centering
   \includegraphics[scale=0.35]{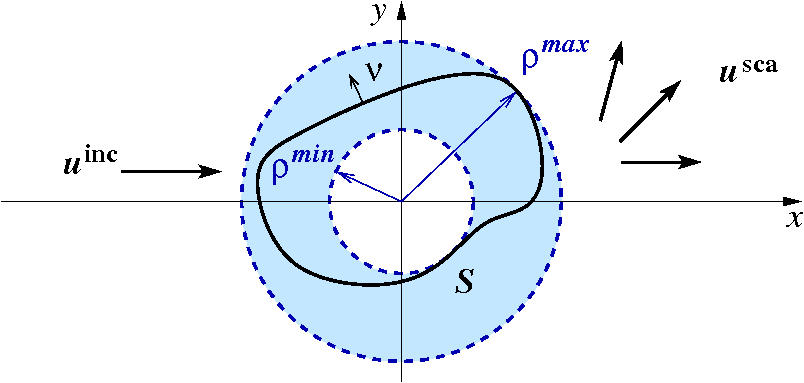}
   \caption{\label{fig:sec4_TransitionRegion}
            \textbf{The Transition Region $\boldsymbol{\Rt}$.}
           }
 \end{figure}

 Our assumption that the obstacle is star-shaped for any realization of the random vector $\ZZ$,
 ensures that its boundary possesses a polar representation,
 \begin{equation}\label{def:sec4_SurfacePolarForm}
  \Scal(\ZZ)
   =
  \left\{
   \left.
   (\rho(\aq;\ZZ)\cdot\cos\aq,\rho(\aq;\ZZ)\cdot\sin\aq)
   \right|
   \,\aq\in[0,2\pi]
  \right\}
   \,,
 \end{equation}
 and the existence of two positive radial bounds, $\rmx$ and $\rmn$, satisfying
 \begin{equation}\label{def:sec4_RadialBounds}
  0
   <
  \rmn
   =
  \inf_{\aq,\zz}\rho(\aq;\zz)
   <
  \sup_{\aq,\zz}\rho(\aq;\zz)
   =
  \rmx
   <
  \infty
   \,.
 \end{equation}
 Thus, as illustrated in Figure \ref{fig:sec4_TransitionRegion}, 
 $\SZ$ is confined to the transition region,
 \begin{equation}\label{def:sec4_TransitionRegion}
  \Rt =
  \left\{
   \left.\rr\in\RR^2\;\right|\;\rmn<\left|\rr\right|<\rmx
  \right\}
   \subset
  \RR^2
  \,.
 \end{equation}
 
 \subsection{Random Shape and Generalized Polynomial Chaos Expansion}\label{sec:SubSection4_2}

 Given our assumptions we observe that the scattering coefficients
 (\ref{def:sec2_SoftScatterCoefInnerProduct2D})
 are finite dimensional \emph{random fields},
 \[
  b_m=b_m(\ZZ)
   \,.
 \]
 A common method to approximate these fields is 
 to obtain their generalized Polynomial Chaos (gPC) expansions,
 \begin{equation}\label{def:sec4_gPCExpansion}
  b_m(\ZZ)\approx b_m^N(\ZZ)
   =
  \sum_{\left|\bn\right|\leq{N}}
  b_{m,\bn}{P}_{\bn}(\ZZ)
   \,,
 \end{equation}
 where $\bn=(n_1,\ldots,n_P)$ is a multi-index and
 $\{{P}_{\bn}\}$ is an orthogonal basis of the inner-product space 
 induced by the probability density function of $\ZZ$,
 \[
  \left< \phi(\ZZ)\,,\,\psi(\ZZ)\right>_{\LG}
   =
  \int_{\IZ}\phi(\zz)\psi(\zz)\rm{d}F_{\ZZ}(\zz)
   \,, \;\;\;
  \rm{d}F_{\ZZ}(\zz)
   =
  \prod_{k=1}^{P}\rm{d}F_{\Zp}(\zp)
   \,,
 \]
 whose support is $\IZ=\Ical_{\Za}\times\cdots\times{\Ical_{\ZP}}$.
 Hence, the expansion coefficients are readily available 
 by the orthogonality via
 \[
  b_{m,\bn}=\frac{1}{\gamma_{\bn}}\left< b_m\,,\,{P}_{\bn}\right>_{\LG}
   \,, \;\;\;
  \gamma_{\bn}=\left< {P}_{\bn}\,,\,{P}_{\bn}\right>_{\LG}
   \,.
 \]

 For a randomly shaped obstacle each coefficient in the gPC expansion
 (\ref{def:sec4_gPCExpansion}) is a target functional of the following form
 \begin{equation}\label{def:sec4_gPCCoefNaive}
  b_{m,\bn}
   =
  \int_{\IZ}
  \int_{\Scal(\zz)}
   g_m(\rr)h(\zz,\ro)\dd\Scal(\zz)
   {{P}_{\bn}(\zz)}\dd F_{\ZZ}(\zz)
  \,.
 \end{equation}
 Using the polar form representation 
 (\ref{def:sec4_SurfacePolarForm}) whose associated 
 induced volume form on the surface is
 $\dd\Scal(\aq;\zz)=\sqrt{\rho^2+\rho_{\aq}^2}\dd\aq$,
 the general representation (\ref{def:sec4_gPCCoefNaive}) 
 can be explicitly written as  
 \begin{equation}\label{def:sec4_gPCCoefNaive_Simple}
  b_{m,\bn}
   =
  \int_{\IZ}
  \int_{[0,2\pi]}
   g_m(\ro)h(\zz,\aq)s(\aq;\zz)
   \rho\dd\aq{{P}_{\bn}(\zz)}\dd F_{\ZZ}(\zz)
  \,,
 \end{equation}
 where the normalized metric tensor in polar coordinates
 is given by
 \[
  s(\aq;\zz)
   =
  \frac{1}{\rho(\aq;\zz)}
   \cdot
  \frac{\partial S(\aq;\zz)}{\partial{\aq}}
   =
  \sqrt{1+\left(\frac{1}{\rho}\cdot\frac{\partial\rho}{\partial\aq}\right)^2}
   \,.
 \]

 In principle, we need to devise a discretization scheme for (\ref{def:sec4_gPCCoefNaive_Simple})
 and apply the optimal reconstruction procedure of Section (\ref{sec:Section3}). However, $\ro\in\SZ$ 
 inherits any irregularity of family of surfaces; e.g., lack of smoothness and oscillatory behaviour,
 which often necessitates specialized high-order discretization of the surface $\SZ$. Additionally,
 discretizing the random surface integral with a grid of numerical integration nodes has to be realized 
 for every grid point in the parameters domain, 
 $\IZ$. Hence, in general, the practical implementation of an inner-product preserving discretization satisfying 
 (\ref{def:sec3_InnerProductPreserve}) is a difficult task. An analytic approach
 for overcoming this fundamental difficulty is presented in the next subsection.

 \subsection{Random Surface Embedding and the Coarea Formula\label{sec:SubSection4_3}}

 In this subsection we present an analytic approach for producing
 inner-product preserving discretizations of functionals of the form of
 (\ref{def:sec4_gPCCoefNaive_Simple}). 
 The key idea is to apply a change of variables transforming 
 (\ref{def:sec4_gPCCoefNaive_Simple})
 to the following equivalent representation
 \begin{equation}\label{def:sec4_gPCCoef_Transformed}
  b_{m,\bn}
   =
  \iint_{\Rt}
   g_m(\rr)
   \phi_{\bn}(h)
  \omega(\rr)r\dd\aq\dd{r}
   \,,
 \end{equation}
 where the weight function $\omega(\rr)>0$, is proportional to
 the conditional expectation of $s(\aq;z)$ given the information
 $\ro(\aq;z)=\rr\in\Rt$. Thus, $\omega(\rr)$ has minimal variance
 while, essentially, encompassing the irregularities of the family of random surfaces, $\SZ$.
 The term $\phi_{\bn}(h)$ is a linear functional uniformly bounded in $\Rt$
 operating on $h$. A high-order numerical integration rule
 with respect to $\omega(\rr)$ would serve as a discretization satisfying 
 (\ref{def:sec3_InnerProductPreserve}). The transformed representation 
 (\ref{def:sec4_gPCCoef_Transformed}) is obtained by the so-called 
 \emph{Coarea Formula} \cite{Federer} which allows us to express the surface
 integral in terms of the integral of the level sets of another function.

 \begin{theorem}{(\bf The Coarea Formula)}\\
  Let $\Dcal$ be an open \emph{Jordan measurable} subset of
  $\RR^{d+\delta d}$ where $d\in\NN$ and $\delta d$ is a non-negative integer.
  Let ${\boldsymbol{\Phi}}:\RR^{d+\delta d}\rightarrow\RR^d$ be a piecewise smooth
  \emph{Lipschitz} function, such that the level set,
  \[
   {\boldsymbol{\Phi}}^{-1}(\rr)
    =
   \left\{
     \right.
    \xx\in\Dcal
     \left|
    {\boldsymbol{\Phi}}(\xx)=\rr
   \right\}
    \,,
  \]
  is a piecewise smooth $\delta d$-dimensional manifold in 
  $\Dcal\subset\RR^{d+\delta d}$.
  Then for any integrable function, $g:\Dcal\rightarrow\RR$,
  we have
  \begin{equation}\label{def:sec4_CoareaFormula}
   \int_{\Dcal}g(\xx){\rm d}\xx
    =
   \int_{\RR^d}
   \left(\int_{{\boldsymbol{\Phi}}^{-1}(\rr)}
    \frac{g(\xx)}{J_{{\boldsymbol{\Phi}}}(\xx)}{\rm d}\Scal_{\rr}(\xx)
   \right){\rm d}\rr
    \,,
  \end{equation}
  where $J_{{\boldsymbol{\Phi}}}(\xx)=\sqrt{\det\left(\left[D_{{\boldsymbol{\Phi}}}(\xx)\right]^T\cdot\left[D_{{\boldsymbol{\Phi}}}(\xx)\right]\right)}$
  is the Jacobian of ${\boldsymbol{\Phi}}$, and ${\rm d}\Scal_{\rr}$
  denotes surface measure of ${\boldsymbol{\Phi}}^{-1}(\rr)$. 
 \end{theorem}
 \begin{remark}
  The coarea formula expresses the integral of a function $g$ over $\Dcal$ 
  in terms of the level sets of
  the function ${\boldsymbol{\Phi}}$. The level sets, ${\boldsymbol{\Phi}}^{-1}(y)$, are called
  fibers of the domain $\Dcal$. The formula is a kind of "curvilinear" version of \emph{Fubini's theorem}.
 \end{remark}

 Let us consider the function
 \[
  {\boldsymbol{\Phi}}(\zz,\aq)
   =
  \ro(\aq;\zz)=(\rho(\aq;\zz)\cdot\cos\aq,\rho(\aq;\zz)\cdot\sin\aq)
   \,, \;\;\;
  {\boldsymbol{\Phi}}:\IZ\times[0,2\pi]\rightarrow\Rt
   \,.
 \]
 By direct calculations we obtain
 \[
  D_{{\boldsymbol{\Phi}}}
   =
  \begin{pmatrix}
   \nabla_{\zz}\rho\cos\aq     & \nabla_{\zz}\rho\sin\aq \\
   \partial_{\aq}(\rho\cos\aq) & \partial_{\aq}(\rho\sin\aq)
  \end{pmatrix}^T\in\RR^{2\times(P+1)}
   \,, \;\;\;
  J_{{\boldsymbol{\Phi}}}(\zz,\aq)
   =
  \rho\cdot\left|\nabla_{\zz}\rho\right|
   \,,
 \]
 where
 \[
  \nabla_{\zz}\rho
   =
  \left(\partial_{\za}\rho,\ldots,\partial_{\zP}\rho\right)^T
   \,, \;\;\;
  \left|\nabla_{\zz}\rho\right|^2
   =
  \sum_{p=1}^P\left(\partial_{\zp}\rho\right)^2
   \,.
 \]
 Now, for applying (\ref{def:sec4_CoareaFormula}) 
 on (\ref{def:sec4_gPCCoefNaive_Simple}) with the 
 chosen implicit function, ${\boldsymbol{\Phi}}$,
 we can only consider spatial points $\rr\in\Rt$ 
 whose associated level set,
 \[
  {\boldsymbol{\Phi}}^{-1}(\rr)
   =
  \left\{
   (\zz,\aq)
    \left|\;
   {\boldsymbol{\Phi}}(\zz,\aq)=\rr
    \right.
  \right\}
   \,,
 \]
 contains at least one smooth $(P-1)$-dimensional manifold in 
 $\RR^{P+1}$. Explicitly, these points satisfy
 $J_{{\boldsymbol{\Phi}}}(\zz,\aq)\neq0$ where
 ${\boldsymbol{\Phi}}^{-1}(\rr)$ is non-empty.
 Thus, we obtain the following representation,
 \begin{align}
  b_{m,\bn}
   & =
  \iint_{\Rt}
   g_m(\rr)
  \int_{\Ical_{\ZZ_{\rr}}}
   h(\zz,\aq)P_{\bn}(\zz)
   \frac{s(\aq;\zz)}{\left|\nabla_{\zz}\rho\right|}
   \prod_{p=1}^{P}\frac{\dd F_{\Zp}}{\dd\zp}
  \dd\Sr r\dd\aq\dd{r}
  \label{def:sec4_gPCCoef_Coarea} \\
   & +
  \int_{\Et}
   g_m(\ro)h(\zz,\aq)s(\aq;\zz)
   \rho\dd\aq{{P}_{\bn}(\zz)}\dd F_{\ZZ}(\zz)
  \,, \label{def:sec4_gPCCoef_Coarea_irregular}
 \end{align}
 where the domain of integration of the inner integral 
 in (\ref{def:sec4_gPCCoef_Coarea}) is given by
 \begin{equation}\label{def:sec4_TheFiberSet}
  \Ical_{\ZZ_{\rr}}
   =
  \left\{
   \left.\zz\in\IZ\right|
   \boldsymbol\rho(\aq;\zz)=\rr\in\Rt
    \,, \;\;\;
   \left|\nabla_{\zz}\rho\right|\neq0
  \right\}
   \subset
  \IZ
   \,,
 \end{equation}
 and the domain of integration of (\ref{def:sec4_gPCCoef_Coarea_irregular})
 is defined by the subset of irregular points of the set ${\boldsymbol{\Phi}}^{-1}(\rr)$,
 \begin{equation}\label{def:sec4_TheIrregularSet}
  \Et
   =
  \left\{
   \left.
    (\zz,\aq)\in\IZ\times[0,2\pi]
   \right|
    \boldsymbol\rho(\aq;\zz)=\rr\in\Rt
    \,, \;\;\;
   \left|\nabla_{\zz}\rho\right|=0
  \right\}
   \subseteq
  {\boldsymbol{\Phi}}^{-1}(\rr)
   \,.
 \end{equation}
 Note that $\Et$ or $\Ical_{\ZZ_{\rr}}$ (for certain values of $\rr$)
 can be empty sets, and in that case the associated integral is taken
 to be zero.

 The advantage of the representation 
 (\ref{def:sec4_gPCCoef_Coarea},\ref{def:sec4_gPCCoef_Coarea_irregular}) 
 is that the wave function, $g_m$, in (\ref{def:sec4_gPCCoef_Coarea})
 is no longer composed with the boundary and does not inherit its irregular properties. 
 The subset of irregular points, $\Et$ (\ref{def:sec4_TheIrregularSet}),
 defines portions of the random surface, $\Scal(\ZZ)$ (\ref{def:sec4_SurfacePolarForm}),
 which are independet on $\ZZ$; i.e., non-random, thus reduces to an integral
 of the following general form,
 \[
  \int_{\Et}
   \left[g_m(\ro(\aq))h(\zz,\aq)s(\aq)
   \rho(\aq)\dd\aq\right]{{P}_{\bn}(\zz)}\dd F_{\ZZ}(\zz)
  \,,
 \]
 whose discretization is straitforward.

 A major challenge is to efficiently evaluate the inner integral
 in (\ref{def:sec4_gPCCoef_Coarea}),
 \begin{equation}\label{def:sec4_gPCCoef_Coarea_InnerIntegral}
  \int_{\Ical_{\ZZ_{\rr}}}
   h(\zz,\aq)P_{\bn}(\zz)
   \frac{s(\aq;\zz)}{\left|\nabla_{\zz}\rho\right|}
   \prod_{k=1}^{K}\frac{dF_{\Zp}}{d\zp}
  \dd\Sr
   \,.
 \end{equation}
 This integral can become infinite since $\left|\nabla_{\zz}\rho\right|^{-1}$
 and $\prod_{k=1}^{K}\frac{dF_{\Zp}}{d\zp}$ are, essentially, singular.
 Accordingly, using the following weight function
 \begin{equation}\label{def:sec4_Coarea_Weight}
  \omega(\rr)
   =
  \int_{\Ical_{\ZZ_{\rr}}}
   \frac{s(\aq;\zz)}{\left|\nabla_{\zz}\rho\right|}
   \prod_{k=1}^{K}\frac{dF_{\Zp}}{d\zp}
  \dd\Sr
   \,,
 \end{equation}
 we have that
 \begin{equation}\label{def:sec4_gPCCoef_Transformed_Again}
  b_{m,\bn}
   =
  \iint_{\Rt}
   g_m(\rr)
   \phi_{\bn}(h)\omega(\rr)
  r\dd\aq\dd{r}
   +
  b_{m,\bn}^{\text{irregular}}
   \,,
 \end{equation}
 where $b_{m,\bn}^{\text{irregular}}$ denotes 
 the irregular component (\ref{def:sec4_gPCCoef_Coarea_irregular}) and
 \begin{equation}\label{def:sec4_gPCCoef_Coarea_InnerIntegral_Disingular}
  \phi_{\bn}(h)
   =
  \omega^{-1}(\rr)
   \cdot
  \int_{\Ical_{\ZZ_{\rr}}}
   h(\zz,\aq)P_{\bn}(\zz)
   \frac{s(\aq;\zz)}{\left|\nabla_{\zz}\rho\right|}
   \prod_{k=1}^{K}\frac{dF_{\Zp}}{d\zp}
  \dd\Sr
   \,,
 \end{equation}
 is a linear functional operating on the surface density $h$.
 If $\phi_{\bn}(h)$ is bounded then the choice (\ref{def:sec4_Coarea_Weight})
 implies that (\ref{def:sec4_gPCCoef_Coarea_InnerIntegral}) is 
 effectively desingularized.

 To show that $\phi_{\bn}(h)$ is a bounded linear functional, 
 let us assume for simplicity that $\Et$ (\ref{def:sec4_TheIrregularSet})
 is an empty set. In that case, we we observe
 that (\ref{def:sec4_Coarea_Weight}) is, in fact, proportional to the conditional expectation 
 of $s(\aq;z)$ given $\rho(\aq;\zz)=r$,
 \[
  \omega(\rr)
   =
  \EE\left[\left. s(\aq;\zz)\right|\rho(\aq,\zz)=r\right]
   \cdot
  f_{\rho}(\zz)
   \,,
 \]
 where $f_{\rho}(\zz)$ is the probability density function of $\rho(\aq;\ZZ)$.
 It is well known that conditional expectation is a minimum variance
 predictor as a function of the given information.
 Hence, the choice (\ref{def:sec4_Coarea_Weight})
 implies minimization of oscillatory behaviour of $s(\aq;\zz)$
 as a function of $\rho(\aq;\zz)$. Employing a similar argument we obtain that
 the linear functional (\ref{def:sec4_gPCCoef_Coarea_InnerIntegral_Disingular})
 is, in fact,
 \[
  \phi_{\bn}(h)
   =
  \frac{\EE\left[\left. h(\zz,\aq)P_{\bn}(\zz)s(\aq;\zz)\right|\rho(\aq;\zz)=r\right]}
       {\EE\left[\left. s(\aq;\zz)\right|\rho(\aq;\zz)=r\right]}
   \,.
 \]
 Hence, the \emph{Cauchy-Schwarz} inequality implies that
 \[
  \left|\phi_{\bn}(h)\right|
   \leq
  \left(\max_{\zz\in\IZr}\left|P_{\bn}(\zz)\right|\right)
   \cdot
  \EE\left[\left. \left| h(\zz,\aq)\right|\;\right|\;\rho(\aq;\zz)=r\right]
   \,,
 \]
 which shows that the linear functional, $\phi_{\bn}$, is, indeed, bounded.
 A similar argument can be applied to show that the functional 
 $\phi_{\bn}$ remains bounded when $\Et$ is not an empty set.

 Setting the integration grid for (\ref{def:sec4_gPCCoef_Transformed_Again})
 can be done in two stages. First we obtain a cubature rule with nodes 
 $\{\rr^{(i,j)}\}\subset\Rt$ and corresponding cubature weights
 $\{\omega^{(i,j)}\}$ with respect to the weight function $\omega(\rr)$,
 regardless of $z$. In the second stage we identify the integration
 grid in $\Ical_{\ZZ_{\rr}}$ which corresponds $\{\rr^{(i,j)}\}$ for the 
 evaluation of $\phi_{\bn}(h)$.

 \subsection{Explicit Representation for a Single Random Variable\label{sec:SubSection4_4}}

 Let us consider a simplified case of one-dimensional random vector,
 \[
  \ZZ
   =
  Z
   \in
  \Ical_Z
   \subset
  \RR
   \,.
 \]
 for which ${\dd}F_{\ZZ}=\dd{F}_{Z}$. We also assume for
 brevity, that $\Et$ (\ref{def:sec4_TheIrregularSet}) is an empty set.
 Thus, the target functional 
 (\ref{def:sec4_gPCCoef_Coarea},\ref{def:sec4_gPCCoef_Coarea_irregular})
 reduces to
 \begin{equation}\label{def:sec4_gPCCoefNaive_Simple_1D}
  b_{m,n}
   =
  \int_{\Ical_Z}
  \int_{[0,2\pi]}
   g_m(\ro)h(z,\aq)s(\aq;z)
   \rho\dd\aq{{P}_{n}(z)}\dd F_{Z}(z)
  \,.
 \end{equation}
 We will show that in this case the functional $\phi_{\bn}(h)$
 (\ref{def:sec4_gPCCoef_Coarea_InnerIntegral_Disingular})
 can be explicitly represented. This approach can serve as
 a building block for the case of a general random vector,
 which is discussed in Section (\ref{sec:Section6}).
 The assumption $\Et=\emptyset$ does not imply
 a loss of generality, since the discretization of
 the irregular part (\ref{def:sec4_gPCCoef_Coarea_irregular})
 is carried out directly without applying the coarea formula.

 Our assumptions imply that for any $\rr\in\Rt$ the fiber set $\Ical_{\ZZ_{\rr}}$ 
 (\ref{def:sec4_TheFiberSet}) is either an empty set or composed of a finite set of discrete points;
 i.e., a zero-dimensional sub-surface.
 Thus, we obtain the following explicit representation 
 of (\ref{def:sec4_gPCCoef_Coarea}) for a single random variable,
 \begin{equation}\label{def:sec4_gPCCoef_Coarea_1D}
  \iint_{\Rt}
   g_m(\rr)
  \left[
   \sum_{z_{\rr}\in\IZr}
   h(z_{\rr},\aq) 
   P_n(z_{\rr})
   \frac{s(\aq;z_{\rr})}{\left|\partial_{z}\rho\right|_{z=z_{\rr}}}
   \frac{dF_{Z}}{dz}(z_{\rr})
  \right]
  r\dd\aq\dd r
   \,,
 \end{equation}
 where $\IZr$ is the reduction of $\Ical_{\ZZ_{\rr}}$ to
 a zero dimensional subset of $\Ical_Z$,
 \begin{equation}\label{def:sec4_TheFiberSet_1D}
  \IZr(\rr)
   =
  \left\{z\in \Ical_Z
         \left| \;
          \rho(\aq;z)=r
         \right.
  \right\}
   \,.
 \end{equation}

 To efficiently evaluate (\ref{def:sec4_gPCCoef_Coarea_1D}),
 we employ the \emph{implicit function theorem} \cite{Dini}
 which ensures the following identities
 \[
  1
   =
  \partial_r\rho(\aq;z_{\rr})
   =
  \left[\partial_z\rho(\aq;z)\right]_{z=z_{\rr}}
  \partial_rz_{\rr}
   \,,
 \]
 \[
  \partial_{\aq}\rho(\aq;z_{\rr})
   =
  \left[\partial_z\rho(\aq;z)\right]_{z=z_{\rr}}
  \partial_{\aq}z_{\rr}
   +
  \left[\partial_{\aq}\rho(\aq;z)\right]_{z=z_{\rr}}
   \,.
 \]
 Thus,
 \begin{equation}\label{def:ImplicitDerivatives_1D}
  \left[\partial_z\rho(\aq;z)\right]_{z=z_{\rr}}
   =
  \frac{1}{\partial_rz_{\rr}}
   \,, \;\;\;
  \left[\partial_{\aq}\rho(\aq;z)\right]_{z=z_{\rr}}  
   =
 -\frac{\partial_{\aq}z_{\rr}}{\partial_rz_{\rr}}
   \,,
 \end{equation}
 and (\ref{def:sec4_gPCCoef_Coarea_1D}) can be 
 equivalently represented by
 \begin{align}
  b_{m,n}
   & =
  \iint_{\Rt}
   g_m(\rr)
  \left[
   \sum_{z_{\rr}\in\IZr}
   h(z_{\rr},\aq)
   P_n(z_{\rr})
   s(\aq;z_{\rr})|\partial_r z_{\rr}|\frac{dF_{Z}}{dz}(z_{\rr})
  \right]
  r\dd\aq\dd r
   \,, \nonumber \\ 
   & =
  \iint_{\Rt}
   g_m(\rr)
  \left[
   \sum_{z_{\rr}\in\IZr}
   h(z_{\rr},\aq)
   P_n(z_{\rr})|\nabla z_{\rr}|\frac{dF_{Z}}{dz}(z_{\rr})
  \right]
  r\dd\aq\dd r
   \,, \label{def:sec4_gPCCoef_Coarea_1D_Implicit} 
 \end{align}
 since the gradient of $z_{\rr}$,
 $\nabla z_{\rr}=\hat{x}\partial_xz_{\rr}+\hat{y}\partial_yz_{\rr}=\rn\partial_rz_{\rr}+\qn\frac{1}{r}\partial_{\aq}z_{\rr}$,
 satisfies $|\nabla z_{\rr}|=s(\aq;z_{\rr})|\partial_r z_{\rr}|$.

 By our definitions $s(\aq;\ab)$ is strictly positive.
 Hence, the zeros of $\partial_{z}\rho$ and the singularities of $\frac{dF_{Z}}{dz}$
 define the integration rule in the sense, that we can apply a cubature rule
 whose weight function captures the singular behaviour of 
 $\left|\partial_rz_{\rr}\right|=\left|\partial_{z}\rho\right|_{z=z_{\rr}}^{-1}$
 and $\frac{dF_{Z}}{dz}$.
 Now, employing (\ref{def:sec4_Coarea_Weight}) yields 
 the following weight function
 \begin{equation} \label{def:sec4_Coarea_Weight_1D_Implicit}
  \omega(\rr)
   =
  \sum_{z_{\rr}\in\IZr}
  s(\aq;z_{\rr})|\partial_r z_{\rr}|\frac{dF_{Z}}{dz}(z_{\rr})
   =
  \sum_{z_{\rr}\in\IZr}
  |\nabla z_{\rr}|\frac{dF_{Z}}{dz}(z_{\rr})
   \,,
 \end{equation}
 which reduces (\ref{def:sec4_gPCCoef_Coarea_1D_Implicit}) to
 \[
  b_{m,n}
   =
  \iint_{\Rt}
   g_m(\rr)\phi_n(h)\omega(\rr)
  r\dd\aq\dd r
   \,,
 \]
 where the bounded linear functional (\ref{def:sec4_gPCCoef_Coarea_InnerIntegral_Disingular})
 reduces to
 \begin{align}
  \phi_n(h)
   & =
  \sum_{z_{\rr}\in\IZr}
   h(\aq;z_{\rr})P_n(z_{\rr})
  \frac{|\nabla z_{\rr}|\frac{dF_{Z}}{dz}(z_{\rr})}
  {\sum_{y_{\rr}\in\IZr}|\nabla y_{\rr}|\frac{dF_{Z}}{dz}(y_{\rr})}
   \,, \nonumber \\ 
   & =
  \sum_{z_{\rr}\in\IZr}
   h(\aq;z_{\rr})P_n(z_{\rr})
   \left[
    \sum_{y_{\rr}\in\IZr\setminus\{z_{\rr}\}}|\nabla y_{\rr}|\frac{dF_{Z}}{dz}(y_{\rr})
   \right]^{-1}
   \,. \label{def:sec4_gPCCoef_Coarea_InnerIntegral_Disingular_1D} 
 \end{align}

 Assuming cubature nodes $\{\rr^{(i)}=(r^{(i)}\cos(\aq^{(i)}),r^{(i)}\sin(\aq^{(i)}))\}$ 
 and corresponding weights $\{\omega^{(i)}\}$ have been chosen, 
 we must also identify the value of $z_{\rr}\in\IZr$ at these nodes
 for the evaluation of $\phi_n(h)$ (\ref{def:sec4_gPCCoef_Coarea_InnerIntegral_Disingular_1D}).
 This, essentially, requires the solution of the following convex minimization problem,
 \[
  \left\{z^{(i,j)}\right\}_{j=1}^{J(i)}
   =
  \argmin_{z\in\Ical_Z} \left(\rho(\aq^{(i)};z)-r^{(i)}\right)^2
   \,,
 \]
 where $J(i)\in\NN$ is the number of solutions to the minimization problem
 for the index $i$.
 Thus, in practice we obtain a cubature grid and corresponding weights
 \[
  \left\{\left(\aq^{(i)},z^{(i,j)}\right)\right\}
   \,, \;\;\;
  \left\{\omega^{(i)}\right\}
   \,,
 \]
 respectively, which apply to the original form (\ref{def:sec4_gPCCoefNaive_Simple_1D})
 in the sense that
 \begin{equation}\label{def:sec4_gPCCoefNaive_Simple_1D_Qubature}
  b_{m,n}
   \approx
  \sum_{i}
   g_m(\rr^{(i)})
   \phi_n^{(i)}
  r^{(i)}\omega^{(i)}
   \,,
 \end{equation}
 where $\rr^{(i)}=\ro(\aq^{(i)},z^{(i,j)})$ independently of $j$,
 and
 \[
  \phi_n^{(i)}
   =
  \sum_{j=1}^{J(i)}
   h(\aq^{(i)};z^{(i,j)})P_n(z^{(i,j)})
  \left[
   \sum_{k\neq j}\left|\nabla z_{\rr}\right|\frac{dF_Z}{dz}
  \right]_{z=z^{(i,k)}}^{-1}
   \,,
 \]
 where $\left|\nabla z_{\rr}\right|_{z=z^{(i,k)}}$
 should, generally, be obtained numerically.

 \subsection{Gaussian Cubature and Error Estimates for a Single Random Variable\label{sec:SubSection4_5}}

 Let us now consider the error estimate of the coefficients $b_{m,n}$ 
 (\ref{def:sec4_gPCCoefNaive_Simple_1D})
 by the numerical cubature (\ref{def:sec4_gPCCoefNaive_Simple_1D_Qubature}).
 As in the previous subsection, we assume for simplicity that $\Et=\emptyset$.
 Since we are interested in representing each coefficient $b_{m,n}$ as
 a weighted \emph{gridfunction} (\ref{def:sec3_WeightedGridfunction}),
 we consider Gaussian iterated quadrature rules whose weights, $\{\omega^{(i)}\}$, 
 are guaranteed to be strictly positive.

 In the literature, error estimates for Gaussian cubature in terms of the integrated
 function derivatives is confined to simple geometries; e.g. circles, spheres and
 convex polygonal shapes. See \cite{Haber1970} for a review.
 Error estimates for more complex shapes can be obtained
 by employing mappings to reference simple shapes. Thus, an exact analysis for the
 problem at hand would be particular to the specific problem and underlying geometry.
 
 Setting a Gaussian cubature in a general $2D$ domain is, typically,
 accomplished by decomposing the domain of integration to subdomains whose interiors do not intersect,
 and applying a distinct cubature in each subdomain.
 This approach is the common practice in spectral methods for partial differential equations
 \cite{boyd2013chebyshev,canuto2012spectral}.
 We assume that the decomposition ensures, that $\omega(\rr)$ is analytic in each subdomain, 
 but possibly singular on the boundary of the subdomain. 
 Thus, mapping each subdomain to a reference simple shape,
 a Gaussian iterated quadrature rule is, essentially, available.

 Assuming each cubature in each subdomain, $\Rcal_q$, of the partition employs $N_{q}$ 
 Gaussian integration nodes, the numerical cubature error in each subdomain 
 is asymptotically $c_{q}\cdot R_s^{-N_{q}}$ for any analytic integrated function in the subdomain.
 The constants $c_{q}$ and $R_q$ are positive, where the latter is a measure of the distance of
 the intervals of integration from the nearest singular point in the complex plane.
 Thus, for sufficiently large $\{N_q\}$ the overall error is satisfies
 \[
  \left|
  b_{m,n}
   -
  \sum_{i}
   g_m(\rr^{(i)})
   \phi_n^{(i)}
  r^{(i)}\omega^{(i)}
  \right|
   \lessapprox
  \sum_qc_qR_q^{-N_q}
   \,,
 \]
 which ensures, asymptotically, exponential convergence.

 \subsection{Example: Randomly Oriented Elliptic Cylinder\label{sec:SubSection4_6}}

\begin{figure}[!h]
  \centering
   \includegraphics[scale=0.35]{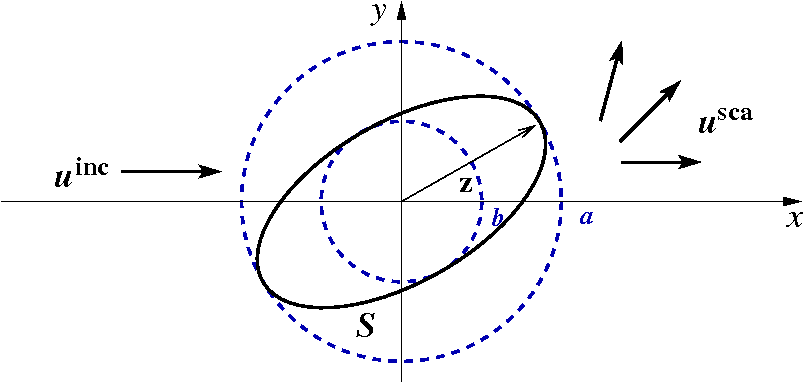}
   \caption{\label{fig:sec4_EllipticCylinder}
    \textbf{Randomly Oriented Elliptic Cylinder.}
   }
 \end{figure}

 Let $\Bcal$ be a sound-soft elliptic cylinder with major radius $a$
 and minor radius $b$, i.e., $a>b>0$. The symmetry axis of the cylinder is
 located at the origin $\Ocal=(0,0)$.
 The major and minor axes of the elliptic cross-section are assumed
 to be rotated counter-clockwise by $\ab\in[0,2\pi]$,
 see Figure \ref{fig:sec4_EllipticCylinder}.
 Note, that the radial bounds (\ref{def:sec4_RadialBounds})
 are $b=\rmn<\rmx=a$. 

 The polar form of the obstacle's boundary over all random orientation states
 is given by
 \[
  \Sz
   =
  \left\{
   \left.\left(\rho(\aq-\ab)\cdot\cos\aq,\rho(\aq-\ab)\cdot\sin\aq\right)\right|
   \,\aq\in[0,2\pi]
  \right\}
   \,,
 \]
 where $\rho(t)\in(b,a)$ satisfies 
 \[
  \rho(t)
   =
  \frac{ab}{\sqrt{b^2\cos^2t+a^2\sin^2t}}
   \,, \;\;\;
  \frac{\rho'(t)}{\rho(t)}
   =
  \frac{(a^2-b^2)\sin(2t)/2}{b^2\cos^2t+a^2\sin^2t}
   \,.
 \]
 Thus, the zeros of $\partial_{\ab}\rho(\aq-\ab)$ are attained at
 $\aq-\ab = 0,\pi/2,\pi,3\pi/2$. Note, that
 \[
  \left[\rho(\aq-\ab)\right]_{\aq-\ab=0,\pi}
   =
  a
   \,, \;\;\;
  \left[\rho(\aq-\ab)\right]_{\aq-\ab=\pi/2,3\pi/2}
   =
  b
   \,,
 \]
 which are, indeed, the points of the random surface
 that do not vary in the radial direction as a function
 of the parameter $z$.

 The equality $\rho(t)=r$ can be solved analytically which yields
 \begin{equation}\label{def:sec4_EllipticCylinderFiberSet}
  \left\{
   \rho(t)=r
    \,|\,
   r\in(b,a)
  \right\}
   =
  \{-\tr\,,\tr\,,\pi-\tr\,,-\pi+\tr\}
   \,,
 \end{equation}
 where
 \begin{equation}\label{def:sec4_EllipticCylinderIntersectionAngle}
  \tr(r)
   = 
  \arccos\left(\frac{a}{r}\sqrt{\frac{r^2-b^2}{a^2-b^2}}\right)
   \in
  (0,\pi/2)
   \,, \;\;\;
  \forall\,r\in(b,a)
   \,.
 \end{equation}
 Hence, we obtain $\IZr=\{\ab_{\rr}^{(1)} , \ab_{\rr}^{(2)} , \ab_{\rr}^{(3)} , \ab_{\rr}^{(4)}\}$
 (\ref{def:sec4_TheFiberSet_1D}) where
 \[
  \ab_{\rr}^{(1)}=\aq-\tr
   \,, \;\;\;
  \ab_{\rr}^{(2)}=\aq+\tr
   \,, \;\;\;
  \ab_{\rr}^{(3)}=\aq-\tr+\pi
   \,, \;\;\;
  \ab_{\rr}^{(4)}=\aq+\tr-\pi
   \,,
 \]
 and
 \[
  \frac{\partial\ab_{\rr}^{(k)}}{\partial r}
   =
  (-1)^p\frac{d\tr}{dr}
   =
  (-1)^{k+1}\left|\frac{d\tr}{dr}\right|
   \,, \;\;\;
  k=1,2,3,4
   \,.
 \]
 Thus, the target functional (\ref{def:sec4_gPCCoef_Coarea_1D_Implicit}) 
 takes the following form
 \[
  b_{m,n}
   =
  \iint_{\Rt}
   g_m(\rr)
  \left[
   \sum_{k=1}^{4}
   h(\aq,z_{\rr}^{(k)})
   P_n(z_{\rr}^{(k)})
   \sqrt{1+\left(\frac{d\tr}{dr}\right)^2}
   \frac{dF_{Z}}{dz}(z_{\rr}^{(k)})
  \right]
  r\dd\aq\dd r
   \,,
 \]
 where
 \[
  \Rt =
  \left\{
   \left.\rr\in\RR^2\;\right|\;b<\left|\rr\right|<a
  \right\}
  \,.
 \]

 Let us now assume that $Z$ is a uniformly distributed random variable in $[0,2\pi]$,
 \[
  \PP(Z<\ab_0)
   =
  \int_{0}^{\ab_0}\frac{\rm{d}\ab}{2\pi}
   =
  \frac{\ab_0}{2\pi}
   \,.
 \]
 The problem is $2\pi$-periodically smooth in $z$, 
 hence, it is natural to employ 
 \[
  P_0(\ab)=1
   \,, \;\;\;
  P_n(\ab) 
   = 
  \left\{\begin{array}{rcl}
   \cos(\lfloor n/2\rfloor \aq) & \text{if} & n   =  2\lfloor n/2\rfloor \\
   \sin(\lfloor n/2\rfloor \aq) & \text{if} & n \neq 2\lfloor n/2\rfloor
  \end{array}\right.
   \;\;\;
  n=1,2,\ldots
   \,,
 \]
 in the gPC expansion (\ref{def:sec4_gPCExpansion}),
 which leads to the following representations
 of the target functional 
 (\ref{def:sec4_gPCCoef_Coarea_1D_Implicit})
 \begin{equation}\label{def:sec4_gPCCoef_Coarea_1D_Implicit_Ellipse}
  b_{m,n}
   =
  \int_{b}^{a}
  \int_{0}^{2\pi}
  g_m(\rr)\phi_n(h)
  \omega(r)
  r\rm{d}\aq\rm{d}r
   \,,
 \end{equation}
 where the linear functional is
 \[
  \phi_n(h)
   =
  \sum_{k=1}^{4}
  h(\aq,\ab_{\rr}^{(k)})
  P_n(\ab_{\rr}^{(k)})
   \,.
 \]
 The weight function 
 (\ref{def:sec4_Coarea_Weight_1D_Implicit})
 is explicitly given by
 \begin{equation}\label{def:sec4_EllipticCylinderWeight}
  \omega(r)
   =
  \sum_{k=1}^{4}
  \frac{1}{2\pi}
  \sqrt{1+\left(\frac{d\tr}{dr}\right)^2} \\
   =
  \frac{\omega^{\text{reg}}(r)}{\sqrt{(a-r)(r-b)}}
  \,,
 \end{equation}
 where the regular part of (\ref{def:sec4_EllipticCylinderWeight})
 is given by
 \[
  \omega^{\text{reg}}(r)
   =
 \frac{2}{\pi}
 \sqrt{\frac{1}{r^2(a+r)(r+b)}+(a-r)(r-b)}
  \,.
 \]

 \subsection{Simulation: Randomly Oriented Elliptic Cylinder\label{sec:SubSection4_7}}

 In this subsection we explore numerically the randomly oriented elliptic cylinder
 example, that was introduced in the previous subsection. First, let us setup
 an inner-product preserving discretization of (\ref{def:sec4_gPCCoef_Coarea_1D_Implicit_Ellipse}).
 Applying the linear change of variables on the radial variable, $r(\sigma)=\frac{a-b}{2}\sigma+\frac{a+b}{2}$,
 which maps $(b,a)$ onto $(-1,1)$, we obtain
 \[
  \frac{1}{\sqrt{(a-r)(r-b)}}
   =
  \frac{2}{a-b}\frac{1}{\sqrt{1-\sigma^2}}
   \,, \;\;\;
  \sigma\in(-1,1)
   \,.
 \]
 Hence, we can employ the Chebyshev-Gauss quadrature in terms of $\sigma$
 for the integration in the radial direction,
 \[
  r^{(i)}
   =
  \frac{a-b}{2}\sigma^{(i)}
   +
  \frac{a+b}{2}
   \,, \;\;\;
  \sigma^{(i)}
   =
  \cos\left(\frac{2i-1}{2M}\pi\right)
   \;\;\;
  i=1,2,\ldots,M
   \,.
 \]
 For the angular variable we employ the composite trapezoidal rule,
 \[
  \aq^{(i,j)} = 2\frac{j-1}{N^{(i)}}\pi
   \;\;\;
  j=1,2,\ldots,N^{(i)}
   =
  10\lfloor r^{(i)}\rfloor
   \,,
 \]
 where $N^{(i)}$ is proportional to $r^{(i)}$ to accommodate for
 the integration over the circumference $2\pi r^{(i)}$. 
 The corresponding spatial cubature formula is
 \[
  \int_{\aq=0}^{2\pi}
  \int_{r=a}^{b}
   \frac{f(r,\aq)\,\dd{r}}{\sqrt{(a-r)(r-b)}}
  \frac{\dd{\aq}}{2\pi}
   \approx
  \sum_{i=1}^{M}
  \sum_{j=1}^{N^{(i)}}
   f(r^{(i)},\aq^{(i,j)})
    \cdot
   \frac{2\pi}{M\cdot N^{(i)}}
  \,.
 \]
 Finally, using (\ref{def:sec4_EllipticCylinderFiberSet}) and
 (\ref{def:sec4_EllipticCylinderIntersectionAngle}) we obtain the following
 expression for the corresponding cubature points in terms of $z$,
 \begin{align}
  z^{(i,j,1)}=\aq^{(j)}-\tr^{(i)} \quad\;\;\; \,,
   & \;\;\;
  z^{(i,j,2)}=\aq^{(j)}+\tr^{(i)} \,,
   \nonumber \\ 
  z^{(i,j,3)}=\aq^{(j)}-\tr^{(i)}+\pi \,,
   & \;\;\;
  z^{(i,j,4)}=\aq^{(j)}+\tr^{(i)}-\pi \,,
   \nonumber 
 \end{align}
 where
 \[
  \tr^{(i)}
   =
  \arccos\left(\frac{a}{r^{(i)}}\sqrt{\frac{r^{(i)^2}-b^2}{a^2-b^2}}\right)
   \,.
 \]

\begin{figure}[!h]
  \centering
  \begin{tabular}{cc}
   \includegraphics[scale=0.265]{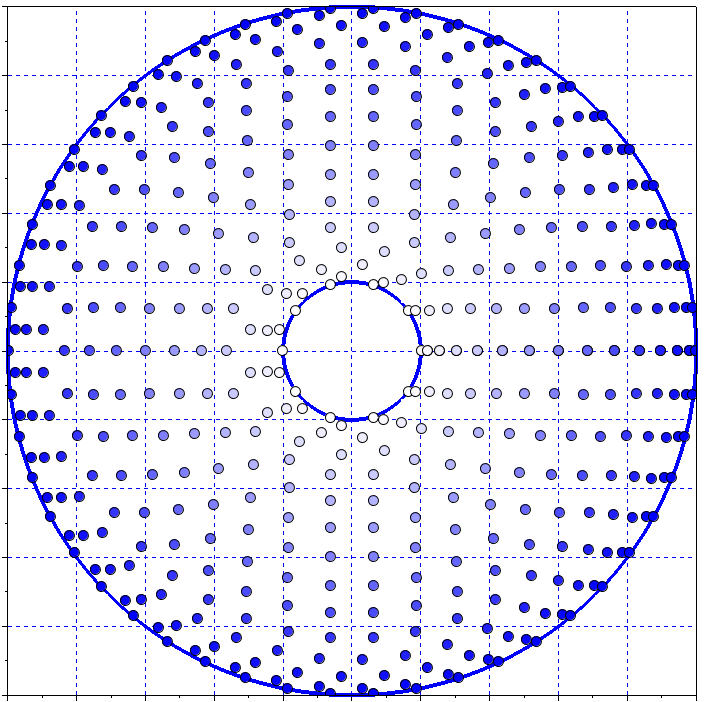} &
   \includegraphics[scale=0.250]{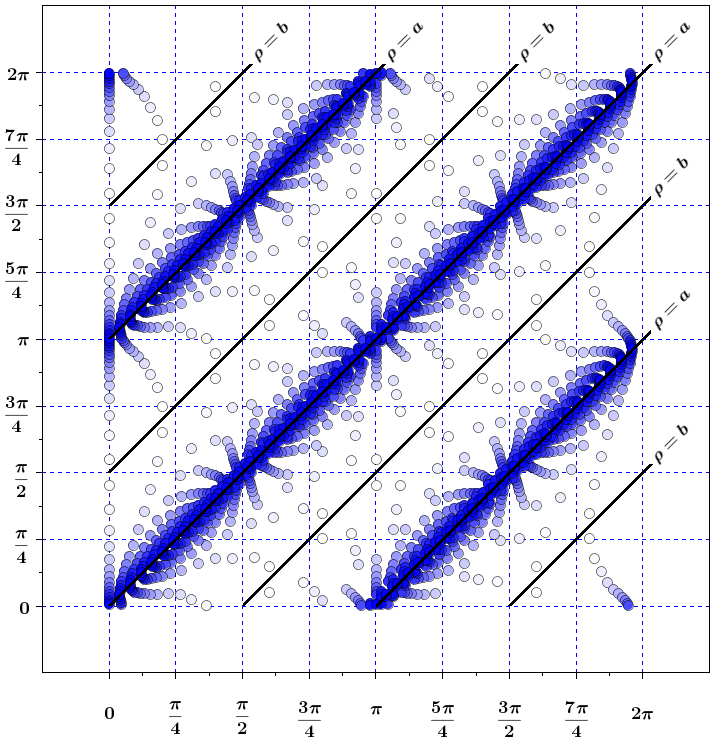} \\
   (a) the annular $(\Rt)$ cubature grid.                               &
   (b) the $(\aq,\ab)$-plane cubature grid.
  \end{tabular}
  \caption{\label{fig:sec4_EllipticCylinderGridCoarea}
   \textbf{Elliptic Cylinder: Coarea discretization.} Displaying the distribution              
   of grid points in the spatial and the parametric domains. Gridpoints intensity               \hfill\textcolor{white}{.}
   shifts from light at $\rmn=b$ to dark at $\rmx=a$. (a) spatial gridpoints in the             \hfill\textcolor{white}{.}
   annulus $\Rt\subset\RR^2$. (b) parametric gridpoints in the $(\aq,\ab)$-plane.               \hfill\textcolor{white}{.}
  }
 \end{figure}

\begin{figure}[!h]
  \centering
  \begin{tabular}{cc}
   \includegraphics[scale=0.265]{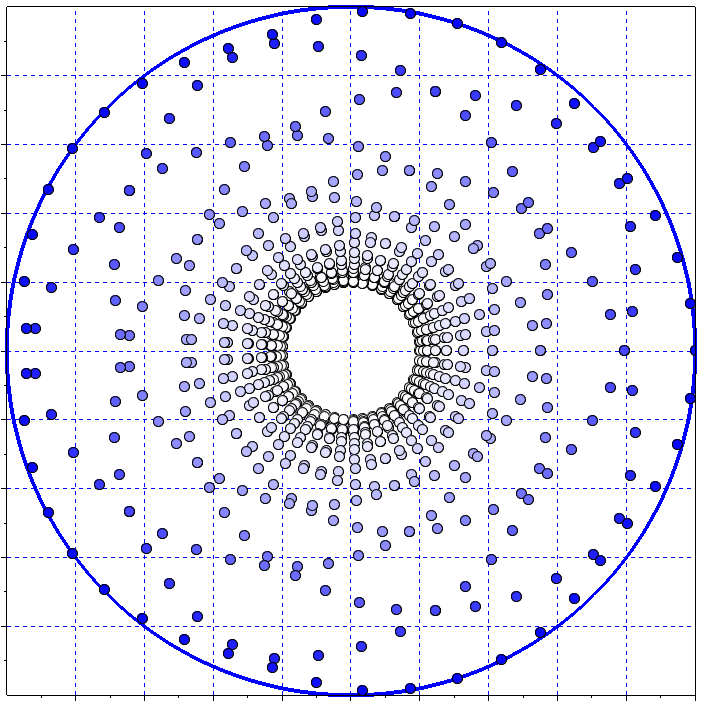} &
   \includegraphics[scale=0.250]{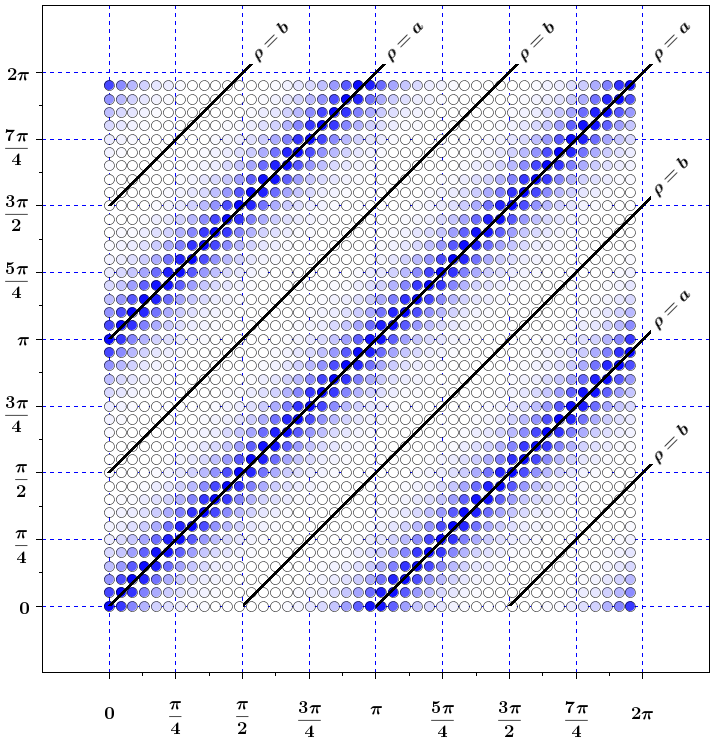} \\
   (a) the annular $(\Rt)$ cubature grid.                               &
   (b) the $(\aq,\ab)$-plane cubature grid.
  \end{tabular}
  \caption{\label{fig:sec4_EllipticCylinderGridNaive}
   \textbf{Elliptic Cylinder: Naive discretization.} Displaying the distribution                \hfill\textcolor{white}{.}
   of grid points in the spatial and the parametric domains. Gridpoints intesity                \hfill\textcolor{white}{.}
   shifts from light at $\rmn=b$ to dark at $\rmx=a$. (a) spatial gridpoints in the             \hfill\textcolor{white}{.}
   annulus $\Rt\subset\RR^2$. (b) parametric gridpoints in the $(\aq,\ab)$-plane.               \hfill\textcolor{white}{.}
  }
 \end{figure}

 The resulting spatial grid $\{(r^{(i)}\cos\aq^{(i,j)},r^{(i)}\sin\aq^{(i,j)})\}$
 in $\Rt$ and the corresponding parametric grid $\{(\aq^{(i,j)},z^{(i,j,k)})\}$ in the
 $(\aq,z)$-plane are displayed in Figure \ref{fig:sec4_EllipticCylinderGridCoarea}.
 For comparison a naive discretization of the random surface, whose parametric grid is 
 uniformly distributed in the $(\aq,z)$-plane,
 \begin{equation}\label{def:sec4_EllipticCylinderGridNaive}
  \aq_{\text{naive}}^{(j)} = 2\frac{j-1}{N}\pi
   \,, \;
  j=1,2,\ldots,N
   \,, \;\;\;
  \ab_{\text{naive}}^{(i)} = 2\frac{i-1}{M}\pi
   \,, \;
  i=1,2,\ldots,M
   \,,
 \end{equation}
 is given in Figure \ref{fig:sec4_EllipticCylinderGridNaive}.
 Evidently, the corresponding naive spatial grid does a poor job in
 properly covering the transition region, $\Rt$.

\begin{figure}[!h]
  \centering
   \includegraphics[scale=0.25]{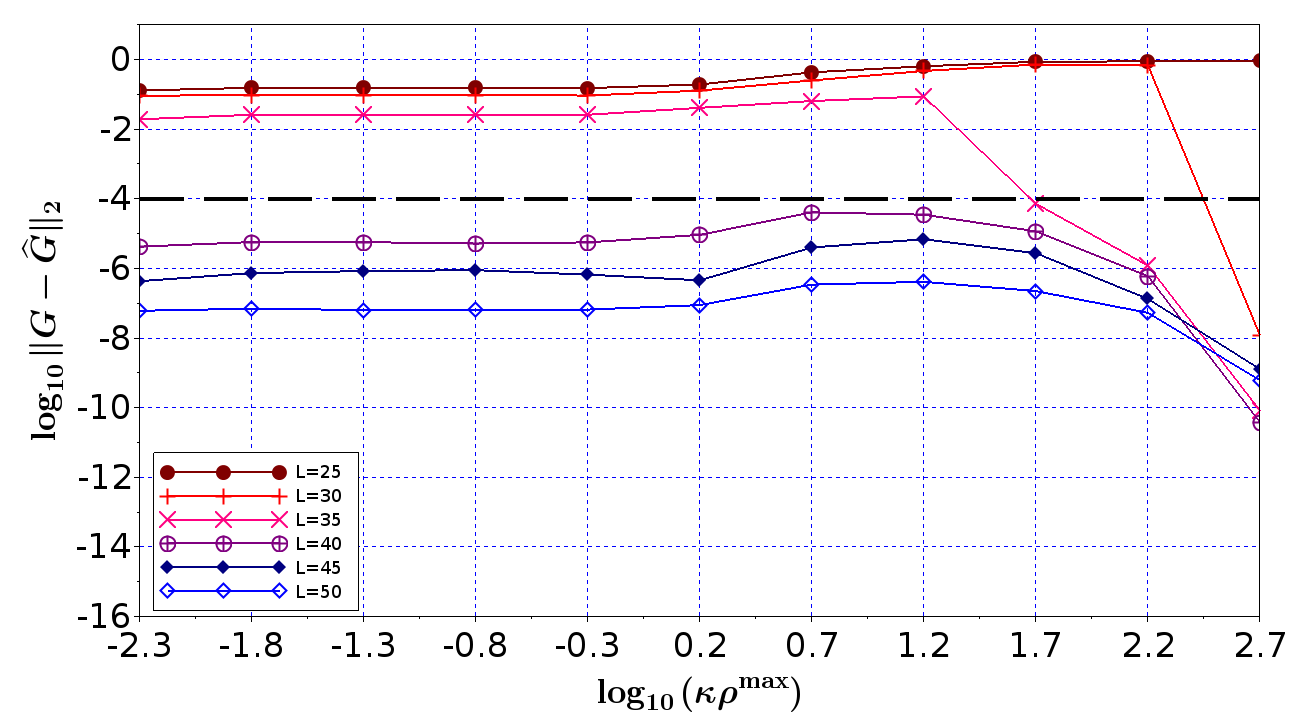}
   \caption{\label{fig:sec4_EllipticCylinderReconstructionError}
    \textbf{Elliptic Cylinder: Reconstruction Error.} Displaying the                                    
    reconstruction error $\|G-\widehat{G}\|$ vs. $\log_{10}(\kk)\rmx$ for various values of the         
    parameter defining the number of target  functionals, $L=25,30,\ldots,50$.                          
   }
 \end{figure}

 For the simulation we consider an elliptic cylinder
 whose semi-major axis is $a=5$ and whose semi-minor 
 axis is $b=1$. We assume an incident plane-wave,
 \begin{equation}\label{def:sec4_EllipticCylinderIncidentField}
  {\uinc}
   =
  e^{\ii\kk x}
   =
  e^{\ii\kk r\cos\aq}
   =
  \sum_{m=-\infty}^{\infty}
  \ii^m\Jm(\kk r)e^{\ii m\aq}
  \,,
 \end{equation}
 which is approximated by truncating the infinite sum in 
 (\ref{def:sec4_EllipticCylinderIncidentField}) to
 a finite sum over the modes $|m|\leq\mu$ where
 $\mu$ satisfies (\ref{def:sec2_BesselSeriesTruncationParameter}).
 For the discretization we have used $M=15$ and $N=10$, and the
 following thresholds for the reconstruction
 \[
  \eev=10^{-4}
   \,, \;\;\;
  \eed=10^{-8}
   \,.
 \]
 Figure \ref{fig:sec4_EllipticCylinderReconstructionError}
 displays the construction error $\|G-\widehat{G}\|_2$,
 where $G$ is a matrix whose rows are the discretized target functionals,
 \[
  g_m(\rr)
   =
  J_m(\kappa|\rr|)e^{-\ii m\aq} P_0(\zz)
   =
  J_m(\kappa|\rr|)e^{-\ii m\aq}
   \,, \;\;\
  |m|\leq\mu
   \,,
 \]
 and $\widehat{G}$ is a matrix whose rows are the corresponding
 reconstructed target functionals from the following set of information functionals
 \[
  f_{\ell,m}
   =
  \Hm(\kappa\left|\rr-\rl\right|)
  e^{\ii m\sphericalangle\left(\rr-\rl\right)}
   \,, \;\;\;
  |m|\leq M
   \,.
 \]
 The singular points $\rl\in\Bcal(Z)$ are uniformly distributed 
 along the family of random surfaces, $\Scal(Z)$,
 \[
  \rl(Z)
   =
  0.95\rho(\al;\ab)\cdot(\cos\al,\sin\al)
   \,, \;\;\;
  \al=2\frac{\ell-1}{L}\pi
   \,, \;\;\;
  \ell=1,2,\ldots,L
   \,.
 \]

\section{Randomly Shaped Polygonal Cylinders}\label{sec:Section5}

 In this section we consider the application of the theory to
 a class of randomly shaped polygonal cylinders.
 This class is characterized by non-smooth randomly varying geometry,
 and thus serves a proof of concept that the proposed method can, indeed, 
 be applied to complex shapes.
 The generalizations to more complex geometries
 is discussed in Section (\ref{sec:Section6}).

 We begin with the introduction of a piecewise smooth polar form
 representation followed by a detailed discussion on the considerations
 for setting the spatial cubature for a single random variable.
 A numerical example including a comparative study with a
 Monte Carlo Nystr\"{o}m approximation concludes this section.
 
 Despite seemingly simplistic at first glance, high-order approximation of 
 wave scattering by a polygonal cylinder is a non-trivial problem.
 The main difficulty stems from the singular behavior of the solution 
 at the corners.
 Development of efficient Nystr\"{o}m discretization techniques for such problems
 have been proposed in recent years \cite{Bremer2012,hao2014high}, 
 and is still an active research topic.
 These techniques, essentially, rely on quadrature based Gauss-Legendre panels
 due to Kolm-Rokhlin \cite{Kolm_Rokhlin_2001}. 
 Typically, the Kolm-Rokhlin algorithm is quite efficient when the wavenumber
 is in the low and mid-frequency regimes, but becomes inefficient as the wave number
 increases, due to the clustering of the quadrature gridpoints near the corners.

 \subsection{Piecewise Smooth Polar Form Representation}

 We consider a star-shape polygonal cylinder in $2D$ which is given by an ordered set of points,
 \[
  \ro_0=(x_0,y_0)\,,\;\ldots\,,\ro_{Q-1}=(x_{Q-1},y_{Q-1})
   \,,
  \ro_Q=\ro_0
   \,,
 \]
 counter-clockwise distributed in $\RR^2$ satisfying,
 \begin{equation}\label{def:sec5_PolyAngles}
  0
   \leq
  \aq_{q-1}=\arctan(y_{q-1}/x_{q-1})
   <
  \arctan(y_{q}/x_{q})=\aq_{q}
   \leq
  2\pi
 \end{equation}
 for all $q=1,2,\ldots,Q-1$,
 that describe the boundary of the polygon
 where $\Scal_q$ is the line segment connecting $\ro_{q-1}$ and $\ro_{q}$,
 \[
  \Scal=\bigcup_{\ell=1}^Q\Scal_{q}
  \,, \;\;\;
  \Scal_q
   =
  \left\{ 
   \left.\ro_{q-1}+t(\ro_q-\ro_{q-1})\right|
   \,t\in[0,1]
  \right\}
   \,, 
 \]
 for all $q=1,2,\ldots,Q$.
 We also assume that the points $\ro_q$ are smooth functions of a random vector $\ZZ\in\IZ$,
 defining simple open differentiable curves that do not intersect in $2D$ space. 
 In particular each curve does not cross itself.

 For a polar representation we consider an arbitrary line segment $\Scal_q$.
 The following equality, 
 \[
  (\xi,\psi)=\ro=\ro_{q-1}+t(\ro_q-\ro_{q-1})
  \,,
 \]
 is equivalent to the system
 \begin{align}
  \xi(\zz,t)  = \rho\cos\aq & = x_{q-1} + t\cdot(x_{q}-x_{q-1}) \,,   \nonumber \\ 
  \psi(\zz,t) = \rho\sin\aq & = y_{q-1} + t\cdot(y_{q}-y_{q-1}) \;\,. \nonumber  
 \end{align}
 where for every $\zz\in\IZ$, $\aq(\zz,t)\in[\aq_{q-1},\aq_q]$.
 Since $t=t(\aq;\zz)\in[0,1]$, we obtain 
 \[
  t(\aq;\zz)
   =
  \frac{x_{q-1}\sin\aq-y_{q-1}\cos\aq}
       {(y_{q}-y_{q-1})\cos\aq-(x_{q}-x_{q-1})\sin\aq}
   =
  \frac{\rho(\aq;\zz)}
       {\rho_{q}(\zz)}
   \cdot
  \frac{\sin(\aq-\aq_{q-1})}
       {\sin(\aq_{q}-\aq_{q-1})}
   \,,
 \]
 where $\rho_{q}(\zz)=\sqrt{x_{q}^2+y_{q}^2}$,
 which leads to
 \begin{align}
  \rho(\aq;\zz)
   & =
  \frac{\rho_{q}\rho_{q-1}\sin(\aq_{q}-\aq_{q-1})}
       {\rho_{q}\sin(\aq_{q}-\aq)+\rho_{q-1}\sin(\aq-\aq_{q-1})}
   \,, \label{def:sec5_Radial2AngularPoly} \\           
  s(\aq;\zz)
   & =
 \frac{|\ro_{q}-\ro_{q-1}|}
      {\rho_{q}\sin(\aq_{q}-\aq)+\rho_{q-1}\sin(\aq-\aq_{q-1})}
  \,.
 \end{align}
 and
 \begin{equation}\label{def:sec5_Polygonal_Boundary_Derivative}
  \left[
   \frac{1}{\rho}
   \frac{\partial\rho}{\partial\aq}
  \right](\aq;\zz)
   =
  \frac{\rho_{q}\cos(\aq_{q}-\aq)-\rho_{q-1}\cos(\aq-\aq_{q-1})}
       {\rho_{q}\sin(\aq_{q}-\aq)+\rho_{q-1}\sin(\aq-\aq_{q-1})}
  \,.
 \end{equation}
 If we assume that the vertices angles, $\aq_q$, are
 constants; i.e, independent of of $\zz$, we obtain
 \begin{align}
  \nabla_{\zz}\rho(\aq;\zz)
   & =
  \left(
   \frac{\rho(\aq;\zz)}
        {\rho_{q}(\zz)}
  \right)^2
   \;\cdot\;
  \frac{\sin(\aq-\aq_{q-1})}
       {\sin(\aq_{q}-\aq_{q-1})}
  \nabla_{\zz}\rho_{q}(\zz)
  \nonumber \\ 
   & +
  \left(
   \frac{\rho(\aq;\zz)}
        {\rho_{q-1}(\zz)}
  \right)^2
   \cdot
  \frac{\sin(\aq_{q}-\aq)}
       {\sin(\aq_{q}-\aq_{q-1})}
  \nabla_{\zz}\rho_{q-1}(\zz)
   \,,
  \nonumber   
 \end{align}
 and
 \begin{equation}\label{def:sec5_Polygonal_Weight_ConstAng}
  \frac{s(\aq;\zz)}{|\nabla_{\zz}\rho(\aq;\zz)|}
   =
  \frac{|\ro_{q}-\ro_{q-1}|/\rho(\aq;\zz)}
  {
  \left|
   \rho_{q  }\sin(\aq_{q}-\aq  )\nabla_{\zz}\rho_{q-1}
   +
   \rho_{q-1}\sin(\aq-\aq_{q-1})\nabla_{\zz}\rho_{q  }
  \right|
  }
 \end{equation}


 \subsection{Considerations for Setting the Spatial Cubature for a Single Random Variable}
 
 Let $\ZZ=Z\in[-1,1]$ with a given probability density function 
 $\frac{dF(z)}{dz}$, and consider the functional
 (\ref{def:sec4_gPCCoef_Coarea_1D_Implicit}) represented as sum of integrals 
 on the sides of the polygon,
 \begin{align}
  b_{m,n}
   & =
  \sum_{q=1}^{Q}
  \iint_{\Rcal_q}
   g_m(\rr)
  \left[
   \sum_{z_{\rr}\in\IZr,\rr\in\Rcal_q}
   h(\aq;z_{\rr})
   P_n(z_{\rr})
   \left|\nabla z_{\rr}\right|
   \frac{dF_{Z}}{dz}(z_{\rr})
  \right]
  r\dd\aq\dd r
   \,, \nonumber \\ 
   & =
  \sum_{q=1}^{Q}
  \iint_{\Rcal_q}
   g_m(\rr)\phi_n^{(q)}(h)\omega^{(q)}(\rr)
  r\dd\aq\dd r
   \,, \label{def:sec5_gPCCoefCoarea_1D_Implicit_Poly} 
 \end{align}
 where the subdomains of integration are
 \begin{equation}\label{def:sec4_TransitionRegion_SegementQ}
  \Rcal_q
   =
  \left\{\left.\rr\in\Rt
          \right| \;
         \exists z\in \Ical_Z\;:\;\rr\in\Scal_q(z)
  \right\}
   \subset
  \RR^2
   \,.
 \end{equation}
 The corresponding weight functions are
 \[
  \omega^{(q)}(\rr)
   =
  \sum_{z_{\rr}\in\IZr}
  |\nabla z_{\rr}|\frac{dF_{Z}}{dz}(z_{\rr})
   \,, \;\;\;
  \rr\in\Rcal_q
   \,,
 \]
 and the bounded linear functional 
 (\ref{def:sec4_gPCCoef_Coarea_InnerIntegral_Disingular_1D})
 is given by    
 \[
  \phi_n^{(q)}(h)
   =
  \sum_{z_{\rr}\in\IZr}
   h(\aq;z_{\rr})P_n(z_{\rr})
   \left[
    \sum_{y_{\rr}\in\IZr\setminus\{z_{\rr}\}}|\nabla y_{\rr}|\frac{dF_{Z}}{dz}(y_{\rr})
   \right]^{-1}
   \,, \;\;\;
  \rr\in\Rcal_q
   \,.
 \]
 Note, that by our definitions for any $q\neq q'$, $\Rcal_q\cap\Rcal_{q'}$ 
 is a zero measure Jordan set and $\cup_{q}\Rcal_q$ is a subset of $\Rt$ 
 but not equal to $\Rt$ (\ref{def:sec4_TransitionRegion}).

 To define a proper cubature rule we must identify
 the zeros of $\partial_z\rho(\aq;z)$ and their local
 behavior, i.e., \emph{Taylor expansion}. 
 Note, that (\ref{def:ImplicitDerivatives_1D}) implies that
 when $\left|\partial_{\aq}z_{\rr}\right|$ exists 
 (i.e., a finite and real) then any zero of 
 $\partial_z\rho(\aq;z)$ is also a zero of
 $\partial_{\aq}\rho(\aq;z)$, which is easier to compute.
 Indeed, by (\ref{def:sec5_Polygonal_Boundary_Derivative})
 it is sufficient to solve the equalities
 \[
  \rho_{q}\cos(\aq_{q}-\aq)=\rho_{q-1}\cos(\aq-\aq_{q-1})
   \,, \;\;\;
  q=1,2,\ldots,Q
   \,.
 \]
 
 Finally, we note that, in general, $\rho(\aq;z)$ as well as $\partial_z\rho(\aq;z)$ are not smooth
 as functions of $\aq$ across the angles $\aq_q(z)$ (\ref{def:sec5_PolyAngles}).
 Thus, we must construct a separate cubature rule for each integral associated with 
 each subdomain $\Rcal_q$ in (\ref{def:sec5_gPCCoefCoarea_1D_Implicit_Poly}).


 \subsection{Numerical Example}

\begin{figure}[!h]
  \centering
   \includegraphics[scale=0.35]{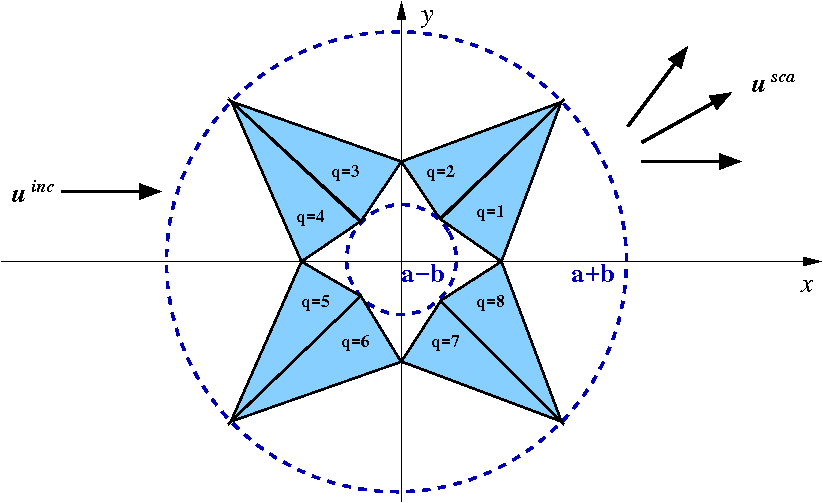}
   \caption{\label{fig:sec5_PolygonalCylinder}
    \textbf{Randomly Shaped Polygonal Cylinder.} 
   }
 \end{figure}

 Consider a star shaped polygonal cylinder whose vertices 
 are defined by 
 \[
  \rho_q
   =
  a+b(1-(-1)^q)\cdot z/2
   \,, \;\;\;
  \aq_q = q\pi/4
   \,, \;\;\;
  q=1,2,\ldots,8
   \,,
 \]
 where $a>b>0$ and $z$ is a realization of a random variable, $Z$,
 uniformly distributed in $[-1,1]$. Note that the radial bounds are
 \[
  \rmn=a-b
   <
  a+b=\rmx
  \,.
 \]
 and the evenly indexed vertices are stationary.
 Thus, induce singularities that have to be dealt carefully,
 as discussed in subsection (\ref{sec:SubSection4_5}).
 An illustration of the regions $\Rcal_q$
 (\ref{def:sec4_TransitionRegion_SegementQ} $(q=1,2,\ldots,8)$) 
 is given in Figure \ref{fig:sec5_PolygonalCylinder}.

 Employing (\ref{def:sec5_Radial2AngularPoly}) we have
 for $\aq_{q-1}\leq\aq\leq\aq_q$,
 \[
  \frac{\rho(\aq;z)}{a}
   =
  \frac{\left(a+bz\right)\cdot\sin(\aq_q-\aq_{q-1})}
       {
        (a+bz)\cdot\sin(\aq_q-\aq)
         +
        a\cdot\sin(\aq-\aq_{q-1})
       }
   \,, \;\;\;
  q=1,3,5,7
   \,,
 \]
 and
 \[
  \frac{\rho(\aq;z)}{a}
   =
  \frac{\left(a+bz\right)\cdot\sin(\aq_q-\aq_{q-1})}
       {
        a\cdot\sin(\aq_q-\aq)
         +
        (a+bz)\cdot\sin(\aq-\aq_{q-1})
       }
   \,, \;\;\;
  q=2,4,6,8
   \,,
 \]
 where $\aq_0=\aq_8(\!\!\!\!\mod2\pi)$. The equality $\rho(\aq;z)=r$
 leads to
 \[
  z_{\rr}
   =
  \frac{a}{b}
  \left[
   \frac{r\sin(\aq-\aq_{q-1})}
        {a\sin(\aq_q-\aq_{q-1})-r\sin(\aq_q-\aq)}
    -
   1
  \right]
   \,, \;\;\;
  q=1,3,5,7
   \,,
 \]
 and
 \[
  z_{\rr}
   =
  \frac{a}{b}
  \left[
   \frac{r\sin(\aq_q-\aq)}
        {a\sin(\aq_q-\aq_{q-1})-r\sin(\aq-\aq_{q-1})}
    -
   1
  \right]
   \,, \;\;\;
  q=2,4,6,8
   \,,
 \]
 where $\rho(\aq;z_{\rr})=r$.

 For obtaining the spatial cubature rule, we 
 employ (\ref{def:sec5_Polygonal_Weight_ConstAng})
 which yields 
 \[
  \frac{s(\aq;z)}{|\partial_{z}\rho(\aq;z)|}
   =
  \frac{|\ro_{q}-\ro_{q-1}|}
       {ab\cdot\rho(\aq;z)}
   \cdot
  \frac{1}{\sin(\aq-\aq_{q-1})}
   \,, \;\;\;
  q=1,3,5,7
   \,,
 \]
 and
 \[
  \frac{s(\aq;z)}{|\partial_{z}\rho(\aq;z)|}
   =
  \frac{|\ro_{q}-\ro_{q-1}|}
       {ab\cdot\rho(\aq;z)}
   \cdot
  \frac{1}{\sin(\aq_q-\aq)}
   \,, \;\;\;
  q=2,4,6,8
   \,.
 \]
 Thus, we seek an efficient cubature rule for each subdomain
 (\ref{def:sec4_TransitionRegion_SegementQ}) approximating
 \begin{equation}\label{def:sec5_gPCCoefCoarea_1D_Poly_Example_Integral}
  \int_{\aq=\aq_{q-1}}^{\aq_q}
  \int_{r=\rho(\aq;-1)}^{\rho(\aq;+1)}
   f_q(r,\aq)
  \dd r \,
   \omega_{q}(\aq)
  \dd\aq
   \,,
 \end{equation}
 where 
 \[
  f_q(r,\aq)
   =
  g_m(\rr)
  h(\aq;z_{\rr})
  P_n(z_{\rr})
  \frac{|\ro_{q}-\ro_{q-1}|}{ab}
 \]
 and the weight function (\ref{def:sec4_Coarea_Weight_1D_Implicit})
 is given by
 \[
  \omega_q(\aq)
   =
  \frac{1}{2}
  \left[
   \frac{1-(-1)^q}{\sin(\aq-\aq_{q-1})}
    +
   \frac{1+(-1)^q}{\sin(\aq_q-\aq)}
  \right]
   \,.
 \]

 Let $\overline{f}_q(\aq)$ denote the mean value of the inner
 integral,
 \[
  \overline{f}_q(\aq)
   \cdot
  \left(
   \rho(\aq;+1)-\rho(\aq;-1)
  \right)
   =
  \int_{r=\rho(\aq;-1)}^{\rho(\aq;+1)}
   f_q(r,\aq)
  \dd r \,
   \,,
 \]
 then (\ref{def:sec5_gPCCoefCoarea_1D_Poly_Example_Integral}) 
 becomes
 \[
  \int_{\aq=\aq_{q-1}}^{\aq_q}
   \overline{f}_q(\aq)
   \left(\rho(\aq;+1)-\rho(\aq;-1)\right)
   \omega_q(\aq)
  \dd\aq
   \,,
 \]
 which is effectively desingularized.
 For obtaining the mean value, $\overline{f}_q(\aq)$,
 we employ the following change of variables,
 \[
   r(\sigma;\aq)
    =
   \frac{\rho(\aq;+1)-\rho(\aq;-1)}{2}
   \sigma
    +
   \frac{\rho(\aq;+1)+\rho(\aq;-1)}{2}
    \,,
 \]
 which yields
 \begin{align}
  \overline{f}_q(\aq)
   & =
  \frac{1}{\rho(\aq;+1)-\rho(\aq;-1)}
  \int_{r=\rho(\aq;-1)}^{\rho(\aq;+1)}
   f_q(r,\aq)
  \dd r 
 \nonumber \\ 
   & =
  2\int_{\sigma=-1}^{1}
   f_q\left(
    \frac{\rho(\aq;+1)-\rho(\aq;-1)}{2}
    \sigma
     +
   \frac{\rho(\aq;+1)+\rho(\aq;-1)}{2}
  \right)
  \dd\sigma
   \,.
 \nonumber
 \end{align}
 Thus, denoting
 \[
  h_q(r,\aq)
   =
 2h(\aq;z_{\rr})
  \frac{|\ro_{q}-\ro_{q-1}|}{ab}
  \left(\rho(\aq;+1)-\rho(\aq;-1)\right)
 \]
 we obtain
 \begin{equation}\label{def:sec5_gPCCoefCoarea_1D_Implicit_Poly_Example}
  b_{m,n}
   =
  \sum_{q=1}^{Q}
  \int_{\aq=\aq_{q-1}}^{\aq_q}
  \int_{\sigma=-1}^1
   g_m(\rr)
   h_q(\rr)
   P_n(z_{\rr})
  \dd\sigma\,
   \omega_q(\aq)
  \dd\aq
 \end{equation}
 where $\rr=\rr(\sigma,\aq)=r(\sigma;\aq)\cdot(\cos\aq,\sin\aq)$.
 Note, that the change of variables in (\ref{def:sec5_gPCCoefCoarea_1D_Implicit_Poly_Example})
 effectively maps each triangular subdomain of integration
 $\Rcal_q$ (\ref{def:sec4_TransitionRegion_SegementQ})
 onto the rectangle $[\aq_{q-1},\aq_q]\times[-1,1]$.

 Accordingly, we propose the following repeated Gauss-Legendre quadrature rules,
 for the approximation of (\ref{def:sec5_gPCCoefCoarea_1D_Implicit_Poly_Example}):
  \begin{enumerate}
   \item Linearly map the angular segment $[\aq_{q-1},\aq_{q}]$ onto $[-1,1]$,
         \[
          \aq(\tau)
           =
          \frac{\pi}{8}\left(\tau+2q-1\right)
           \,, \;\;\;
          \tau\in[-1,1]
           \,.
         \]
   \item Evaluate the $N_q\in\NN$ Gauss-Legendre quadrature nodes in $\tau$,
         \[
          -1<\tau^{(q,1)}<\tau^{(q,2)}<\cdots<\tau^{(q,N_q)}<1
           \,.
         \]
   \item For each $\aq^{(q,j)}=\aq(\tau^{(q,j)})$ apply linear map in the radial direction
         \[
          r^{(q,j)}(\sigma)
           =
          \frac{\rho(\aq^{(q,j)};1)-\rho(\aq^{(q,j)};-1)}{2}\cdot\sigma
           +
          \frac{\rho(\aq^{(q,j)};1)+\rho(\aq^{(q,j)};-1)}{2}
           \,.
         \]
   \item Evaluate the $M_q\in\NN$ Gauss-Legendre quadrature nodes in $\sigma$,
         \[
          -1<\sigma^{(q,1)}<\sigma^{(q,2)}<\cdots<\sigma^{(q,M_q)}<1
           \,.
         \]
  \end{enumerate}

 The overall discretized approximation of 
 (\ref{def:sec5_gPCCoefCoarea_1D_Implicit_Poly_Example})
 becomes
 \[
  \sum_{q=1}^{8}
  \sum_{i=1}^{M_q}
  \sum_{j=1}^{N_q}
   g_m(\rr^{(q,i,j)})
   h_q(\rr^{(q,i,j)})
   P_n(z_{\rr^{(q,i,j)}})
   \omega^{(q,i,j)}
  \,,
 \]
 where
 \begin{align}
  r^{(q,i,j)}
   & =
  \frac{\rho(\aq^{(q,j)};1)-\rho(\aq^{(q,j)};-1)}{2}\cdot\sigma^{(q,i)}
   +
  \frac{\rho(\aq^{(q,j)};1)+\rho(\aq^{(q,j)};-1)}{2}
   \,, \nonumber \\ 
  \rr^{(q,i,j)}
   & = 
  (r^{(q,i,j)}\cos\aq^{(q,j)},r^{(q,i,j)}\sin\aq^{(q,j)})
   \,, \nonumber \\ 
  \omega^{(q,i,j)}
   & =
  \frac{\pi}{8}
  \frac{2}{\left[(1-\sigma^2)P_{N_q}'(\sigma)\right]_{\sigma=\sigma^{(q,i)}}}
   \cdot
  \frac{2}{\left[(1-\tau^2)P_{M_q}'(\tau)\right]_{\tau=\tau^{(q,j)}}}
   \,. \nonumber 
 \end{align}

\begin{figure}[!h]
  \centering
   \includegraphics[scale=0.2]{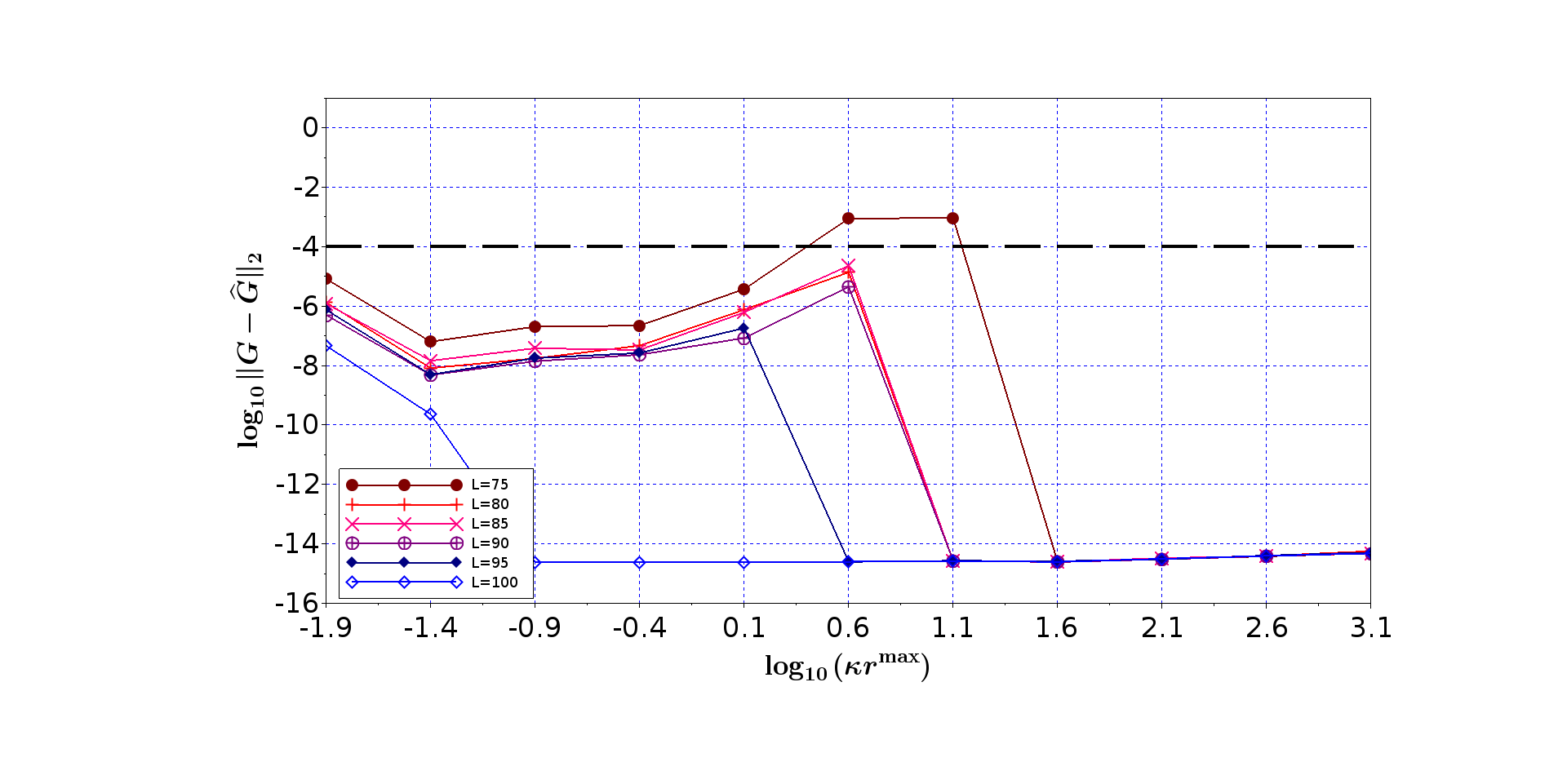}
   \caption{\label{fig:sec5_PolygonalCylinderRErr}
    \textbf{Polygonal Cylinder: Reconstruction Error.} 
    Displaying the reconstruction error $\|G-\widehat{G}\|$ 
    vs. $\log_{10}(\kk)\rmx$ for various values of the  
    parameter defining the number of target  functionals, $L=75,80,\ldots,100$.
   }
 \end{figure}

 For the simulation we set the parameters
 as $a=5$, $b=4$. 
 We assume an incident plane-wave,
 \begin{equation}\label{def:sec5_PolygonalCylinderIncidentField}
  {\uinc}
   =
  e^{\ii\kk x}
   =
  e^{\ii\kk r\cos\aq}
   =
  \sum_{m=-\infty}^{\infty}
  \ii^m\Jm(\kk r)e^{\ii m\aq}
  \,,
 \end{equation}
 which is approximated by truncating the infinite sum in 
 (\ref{def:sec5_PolygonalCylinderIncidentField}) to
 a finite sum over the modes $|m|\leq\mu$ where
 $\mu$ satisfies (\ref{def:sec2_BesselSeriesTruncationParameter}).
 For the discretization we have used $M_q=15$ and $N_q=12$, and the
 thresholds $\eev=10^{-4}$ and $\eed=10^{-8}$ for the reconstruction.
 Figure \ref{fig:sec5_PolygonalCylinderRErr}
 displays the construction error $\|G-\widehat{G}\|_2$,
 where $G$ is a matrix whose rows are the discretized target functionals,
 \[
  g_{m}(\rr)
   =
  J_m(\kappa|\rr|)e^{-\ii m\aq} P_0(\zz)
   =
  J_m(\kappa|\rr|)e^{-\ii m\aq}
   \,, \;\;\
  |m|\leq\mu
   \,,
 \]
 and $\widehat{G}$ is a matrix whose rows are the corresponding
 reconstructed target functionals. For the reconstruction
 We employed the following information functionals
 \[
  f_{\ell,m}
   =
  \Hm(\kappa\left|\rr-\rl\right|)
  e^{\ii m\sphericalangle\left(\rr-\rl\right)}
   \,, \;\;\;
  |m|\leq M_q
 \]
 whose singularities $\rl=0.95\rho(\al;\ab)\cdot(\cos\al,\sin\al)\in\Bcal(Z)$ 
 are uniformly distributed along the family of random surface, $\Scal(Z)$;
 $\al=2\frac{\ell-1}{L}\pi$, $\ell=1,2,\ldots,L$.

\begin{figure}[!h]
  \centering
   \includegraphics[scale=0.2]{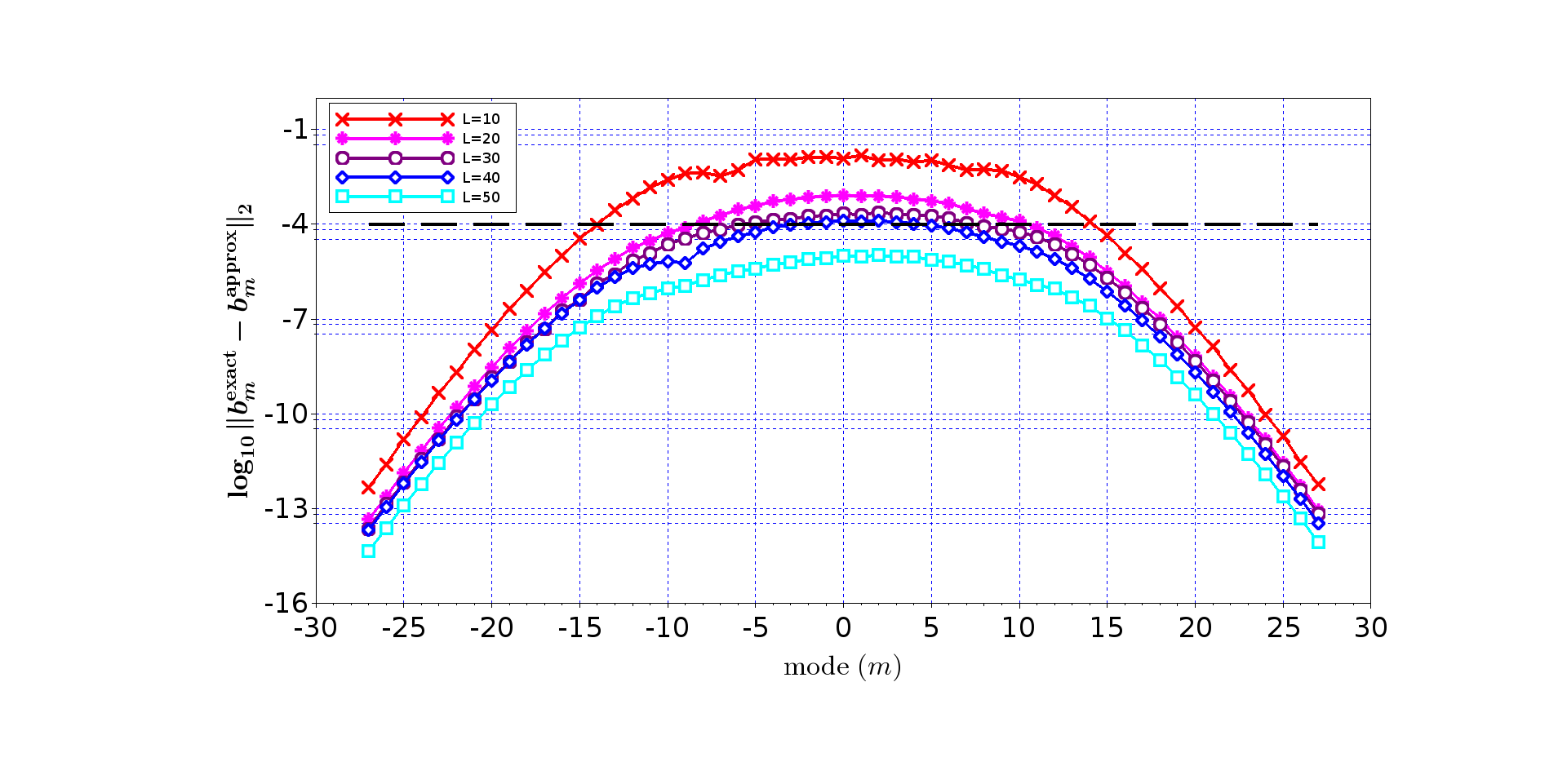}
   \caption{\label{fig:sec5_PolygonalCylinderCErr}
    \textbf{Polygonal Cylinder: Coefficients Error.} 
    Displaying the coefficients expectation error, 
    $|b_m^{exact}-b_m^{approx}|$ in logarithic scale vs. 
    the expansion mode, $m$ for various values of the  
    parameter defining the number of target functionals, $L$.
   }
 \end{figure}

 The results displayed in Figure \ref{fig:sec5_PolygonalCylinderRErr}
 only show the reconstruction error, which may not predict
 the actual error. Hence, to further validate the result,
 the error of the estimated coefficients
 (\ref{def:sec5_gPCCoefCoarea_1D_Implicit_Poly_Example}) of
 the expectation of the scattered wave,
 \[
  b_m^{\text{approx}}
   =
  \sum_{q=1}^{8}
  \sum_{i=1}^{M_q}
  \sum_{j=1}^{N_q}
   g_m(\rr^{(q,i,j)})
   h_q(\rr^{(q,i,j)})
   \omega^{(q,i,j)}
 \]
 compared to the same coefficients, denoted as $b_m^{\text{exact}}$,
 obtained by Monte Carlo simulation with $2000$ uniformly distributed samples of $z$,
 is displayed in Figure \ref{fig:sec5_PolygonalCylinderCErr}.
 Each realization was solved by the 
 Gauss-Legendre paneled Nystr\"{o}m discretization
 using the Kolm-Rokhlin algorithm, where each panel was
 discretized with $15$ quadrature points.
 The comparison was performed for a single wavenumber, $\kappa=1$,
 since the conventional paneled Nystr\"{o}m method is, essentially, unreliable
 for large wavenumbers. From the results it is evident, that in this
 particular example, the actual error is a mgnitude less than the reconstruction
 error. This result is to be expected, since the reconstruction error estimates 
 represent the worst case scenario.


\section{Summary, Conclusions and Future Study}\label{sec:Section6}

 The present paper introduced an alternative approach for quantifying
 the effects of random shape in acoustic scattering problems. The core idea of the 
 proposed method is to construct a spatial embedding within the family of random surfaces,
 which facilitates the construction of a spatial low-dimensional integration rule
 adapted to the underlying random geometry.
 The chosen integration weight function is positive, encompasses
 random surface irregularities and of minimal variance. This, essentially,
 avoids the fundamental problems associated with random surface discretizations,
 namely lack of smoothness in the proximity of the surface when using
 a level-set method and strong non-linear dependence on the variation of the boundary
 when utilizing random domain mapping.
 The method was demonstrated on a pair of model problems in $\RR^2$.

 \subsection{Efficiency and Qualitative Comparison with other Methods}

 Evaluating the full efficiency of the method compared to other techniques
 is a complicated task, especially if one considers parallel implementation.
 However, the main contribution of this work is the analysis and proposed
 framework for constructing numerical integration rules, 
 which minimize the number of integration gridpoints
 required for an accurate evaluation of the solution. Hence, we limit
 the discussion to this aspect.

 First, let us consider null-field reconstruction based on a naive discretization
 which is obtained by sampling the random parameters followed by spatial discretization
 for each realization as demonstrated in \ref{def:sec4_EllipticCylinderGridNaive}.
 Each spatial grid has to be sufficiently accurate to capture the desired functionals
 operating on the sampled surface, which can be complicated. Thus, for complex geometries
 a large number of gridpoints are, generally, required for an accurate approximation.
 Indeed, deterministic null-field reconstruction of wave scattering by elliptic and square cylinders
 as presented in \cite{NFM-Optimal-Reconstruction,NFM-Thesis}, required approximately $500$ gridpoints
 for an efficient approximation. Thus, for a complete reconstruction including the samples
 of the random variables a total of $5,000$ to $10,000$ gridpoints are required.
 Note, that employing stochastic collocation or Monte Carlo, where each realization has to be 
 separately discretized, would lead to similar computational costs.
 Clearly, the coarea discretization, which in our examples, required roughly no more than
 $1,500$ grid points in the polygonal cylinder example, is highly more efficient. 

 The other techniques capable of handling large variations of the random boundary,
 namely level-set and random domain mapping, rely on discretization
 of the spatial domain, whereas the method proposed in this study relies on 
 discretization of the random surface which is of one dimension lower. Thus, 
 the proposed method inherently requires a much smaller discretization grid or,
 equivalently, level of discretization. In addition, the solution obtained
 by the proposed method automatically satisfies the far-field radiation condition,
 whereas the other methods rely on truncation of the spatial domain and some absorbing 
 boundary condition which can reduce the accuracy of the solution.

 The popularity of the random domain mapping method is due to its straightforward nature.
 However, the mapping to the reference domain has to
 be chosen, and can be costly in the case of complex geometry. The mapping typically
 results in highly non-linear coefficients, whose behavior requires a large number
 of gridpoints or level of discretization to capture. In general, this is a brute
 force approach which ignores the particular geometry of the problem and, often,
 requires a high level of discretization which is combined with dimensionality reduction techniques 
 to ensure reasonable computational effort. Also, note that previous studies on random domain
 mapping did not fully consider non-smooth random domains, as presented in this study.

 \subsection{Future Study}

 The current study presented a proof of concept by restricting the analysis
 to $2D$ star shaped obstacles. The major challenges for future work are
 the extension of the analysis to non star shaped $3D$ obstacles, 
 and the full and efficient implementation for more than one random variable. 
 These challenges are discussed in the current subsection.

 Generalizing the new approach to non star-shaped obstacles as well as
 to $3D$ spatial setting is straightforward.
 Indeed, given a non star-shaped obstacle we can
 represent its random boundary as a union of star-shaped sub-surfaces 
 each with a local origin. For a star-shaped obstacle in $\RR^3$ the surface
 of the obstacle possesses a spherical representation,
 \[
  \Scal^{3D}(\ZZ)
   =
  \left\{
   \left.
    (\rho\cos\aq\cos\aj,\rho\cos\aq\sin\aj,\rho\sin\aq)
   \right|
    \,
   \aq\in[0,\pi]
    \,,\; 
   \aj\in[0,2\pi]
  \right\}
   \,,
 \]
 where $\rho=\rho(\aq,\aj;\ZZ)$ corresponds to the spatial vector
 \[
   \rr
    =
   (r\sin\aq\cos\aj,r\sin\aq\sin\aj,r\cos\aq)
    \,.
 \]
 The expansion of the scattered field in $3D$ is of the form
 \begin{equation}\label{sec6_Definition-Scattered-Wave-Analytic-Expansion}
   \usca(\rr)
    = 
   \sum_{n=0}^{\infty}\sum_{m=-n}^{n}b_n^m{\rm h}_n^{(1)}(\kr)Y_n^m\left(\theta,\varphi\right)
    \,,
 \end{equation}
 where  ${\rm h}_n^{(1)}(z)$ is the $n$th-order \emph{spherical Hankel function of the first kind}
 and
 \begin{equation}
  Y_n^m=P_n^m\left(\cos\theta\right)e^{\ii m\aj}
  \,,
 \end{equation}
 where $P_n^m$ are the \emph{associated Legendre functions}.
 See \cite{BIE-CK-InverseScattering-Book} for further details.
 The gPC expansion of the scattering
 coefficients, $b_n^m$, can be represented by
 \[
  b_{n,\bn}^m
   =    \!\!\!\!\!\!\!\!\!\!
  \int\limits_{\qquad\quad\zz\in\IZ}\!\!\!\!\!\!\!\!\!\!\!\cdots\int
  \left[\;
  \int
   \limits_{\aq=0}^{\pi}\int\limits_{\aj=0}^{2\pi}
    h(\ro) \cdot g_n^m(\ro)
    P_{\bn}(\zz)
    S(\aq,\aj;\zz)
   \dd\aq\dd\aj
  \right]
   \prod_{p=1}^{P}\dd F_{\Zp}
  \,,
 \]
 where $g_n^m(\rr)=-\jn(\kappa\rr)P_n^{|m|}(\cos\aj)e^{\iim\aq}$,
 $\ro(\aq,\aj;\zz)=(\rho\cos\aq\cos\aj,\rho\cos\aq\sin\aj,\rho\sin\aq)$
 and the metric tensor for the spherical case is given by
 \[
  S(\aq,\aj;\ZZ)
   =
  \rho
  \sqrt{
   \left(\rho^2+\left(\frac{\partial\rho}{\partial\aj}\right)^2\right)\sin^2\aj
    +
   \left(\frac{\partial\rho}{\partial\aq}\right)^2
  }
   \qquad\quad
  \left(\rho=\rho(\aq,\aj;\zz)\right)
   \,.
 \]
 Employing the Coarea formula with respect to the level sets 
 of ${\boldsymbol{\Phi}}(\zz,\aq,\aj)=\ro(\aq,\aj;\zz)$ yields 
 an expression similar to (\ref{def:sec4_gPCCoef_Coarea})
 \[
  b_{n,\bn}^m
   =
  \iiint\limits_{\Rt}
   g_n^m(\rr)\phi_{\bn}(h)\omega(\rr)
   r^2\sin\aq\dd\aq\dd{r}
  \,,
 \]
 where the $3D$ transition region is
 \[
  \Rt =
  \left\{
   \left.\rr\in\RR^3\;\right|\;\inf_{\aq,\aj,\zz}\rho(\aq,\aj;\zz)<\left|\rr\right|<\sup_{\aq,\aj,\zz}\rho(\aq,\aj;\zz)
  \right\}
   \subset
  \RR^3
  \,,
 \]
 and the weight function is proportional to the conditional expectation
 of the normalized metric tensor,
 \[
  \omega(\rr)
   =
  \EE\left[\left. \frac{S(\aq,\aj;\zz)}{\rho^2\sin\aq}\right|\rho(\aq,\aj;\zz)=r\right]
   \cdot
  f_{\rho}(\zz)
   \,.
 \]
 The expressions are, however, technically more complicated to work with.
 The usage of automatic integration as well as optimization methods for
 obtaining the spatial grid may prove to be a necessity.

 Another challenge is to efficiently deal with several random variables.
 This can be achieved by employing the single random variable formula (\ref{def:sec4_gPCCoef_Coarea_1D})
 as a basic building block.
 Indeed, in the $2D$ case, choosing a single random variable, $Z_{p}$, 
 and applying the coarea formula (\ref{def:sec4_gPCCoef_Coarea}) with respect to the corresponding 
 integration variable $z_{p}$ yields the following equivalent representation
 \[
  b_{m,\bn}
   =
  \iint\limits_{\Rt}
   g_m(\rr) \phi_{\bn}(h) \omega(\rr)
  r\dd\aq\dd{r}
  \,,
 \]
 where the functional operating on $h$ explicitly satisfies
 \[
  \phi_{\bn}(h) \omega(\rr)
   = \!\!\!\!\!\!\!\!\!\!\!\!\!\!\!
  \int\limits_{\qquad\quad\IZ\setminus \Ical_{Z_p}}
   \!\!\!\!\!\!\!\!\!\!\!\!\!\!\!
  \cdots\int\left[
   \sum_{z_{p,\rr}\in\Ical_{Z_{p,\rr}}} \!\!\!\!\!
   h(\zz,\aq) 
   P_{\bn}(\zz)
   \frac{s(\aq;z)}{\left|\partial_{z_p}\rho\right|_{z_p=z_{p,\rr}}}
   \frac{dF_{Z_p}}{dz_p}(z_{p,\rr})
  \right]
  \prod_{q\neq p}\dd F_{Z_q}
   \,,
 \]
 and the zero dimensional fiber set corresponding to $Z_p$ is
 \[
  \Ical_{Z_{p,\rr}}
   =
  \left\{
   \left.z\in\Ical_{Z_p}\right|
   \boldsymbol\rho(\aq;\zz)=\rr
  \right\}
   \subset
  \Ical_{Z_p}
   \,.
 \]
 Note that $z_{p,\rr}$ is a function of $\rr$ as well as $z_1,\ldots,z_{p-1},z_{p+1},\ldots,z_P$.
 In general, globally using the last formula is not expected to produce an optimal result. A more 
 sophisticated approach is to partition $\Rt$ into subregions where each subregion
 is associated with one significant random variable. This, requires performing sensitivity
 analysis, similar to the \emph{analysis of variance} (ANOVA) method \cite{holtz2010sparse}, 
 that is also dependent on the spatial coordinates, $\rr$. The exploration of this
 approach needs a separate extensive study, and deferred to future work.


\bibliographystyle{spmpsci}     
\bibliography{myref_rev}        


\end{document}